\documentclass[a4paper,11pt]{article}
\pdfoutput=1 
\usepackage{jcappub} 
\usepackage[T1]{fontenc}
 \usepackage{amssymb,amsfonts,slashed,amsthm,amsmath,graphicx}
\usepackage[all]{xy}
\usepackage{dcolumn}
\usepackage{bm}
\usepackage{mathrsfs}
\usepackage{ucs}
\usepackage[utf8x]{inputenc}
\usepackage{hyperref}
\usepackage{hepunits}
\usepackage{slashed}
\usepackage{floatrow}
\usepackage{subfigure}
\usepackage{indentfirst}
\usepackage{comment}
\usepackage{multicol}

\begin{document}
\definecolor{green}{rgb}{0.0,0.5,0.0}
\definecolor{purple}{rgb}{0.5,0,0.5}

\newcommand{\Planck}{{\it Planck}}
\newcommand{\LCDM}{$\Lambda$CDM}

\newcommand{\PL}{$\mathcal{PL}$}
\newcommand{\PLr}{$\mathcal{PL}+r$}
\newcommand{\PLgmu}{$\mathcal{PL}+G\mu$}
\newcommand{\PLrgmu}{$\mathcal{PL}+r+G\mu$}

\newcommand{\gmu}{G\mu}

\newcommand{\ie}{\textit{i.e.\ }}
\newcommand{\eg}{\textit{e.g.\ }}

\newcommand{\ora}{\textcolor{orange}}
\newcommand{\bla}{\textcolor{black}}
\newcommand{\blu}{\textcolor{blue}}
\newcommand{\red}{\textcolor{red}}

\newcommand{\be}{\begin{equation}}
\newcommand{\ee}{\end{equation}}
\newcommand{\bq}{\begin{eqnarray}}
\newcommand{\eq}{\end{eqnarray}}
\newcommand{\bsq}{\begin{subequations}}
\newcommand{\esq}{\end{subequations}}
\newcommand{\bc}{\begin{center}}
\newcommand{\ec}{\end{center}}

\newcommand{\xis}{\xi^{\rm s}}
\newcommand{\xim}{\xi^{\rm m}}
\newcommand{\Ls}{\ell_{\rm s}}
\newcommand{\mus}{\mu_{\rm s}}
\newcommand{\musV}{\mu_{V}}
\newcommand{\mumV}{M_{V}}
\newcommand{\betas}{\beta^{\rm s}}
\newcommand{\xiws}{\xi_{\rm w}^{\rm s}}
\newcommand{\xiLs}{\xi_{E}^{\rm s}}
\newcommand{\xiLm}{\xi_{E}^{\rm m}}
\newcommand{\betaws}{\beta_{\rm w}^{\rm s}}
\newcommand{\betaLs}{\beta_{E}^{\rm s}}
\newcommand{\betaLm}{\beta_{E}^{\rm m}}

\bibliographystyle{JHEP}

\subheader{\begin{flushright}
HIP-2017-07/TH
\end{flushright}}

\title{\boldmath {\LARGE{Cosmic Microwave Background constraints for global strings and global monopoles \vspace{-0.3 cm}}}}

\author[a]{{\large Asier Lopez-Eiguren,}}
\author[a,b]{{\large Joanes Lizarraga,}}
\author[c,d]{{\large Mark Hindmarsh}}
\author[a]{{\large and Jon Urrestilla}}

\affiliation[a]{Department of Theoretical Physics, University of the Basque Country UPV/EHU,\\
48080 Bilbao, Spain}
\affiliation[b]{Department of Applied Mathematics, University of the Basque Country UPV/EHU,\\
48013 Bilbao, Spain.}
\affiliation[c]{Department of Physics \& Astronomy, University of Sussex, Brighton, BN1 9QH, United Kingdom}
\affiliation[d]{Department of Physics and Helsinki Institute of Physics, PL 64, FI-00014 University of Helsinki, Finland}

\emailAdd{asier.lopez@ehu.eus}
\emailAdd{joanes.lizarraga@ehu.eus}
\emailAdd{m.b.hindmarsh@sussex.ac.uk}
\emailAdd{jon.urrestilla@ehu.eus}

\abstract{ We present the first
cosmic microwave background (CMB) power spectra from 
numerical simulations of the global O($N$) linear $\sigma$-model, with $N=2,3$, 
which have global strings and monopoles as topological defects.
In order to compute the CMB power spectra we 
compute the unequal time correlators (UETCs) of the energy-momentum tensor, showing that they fall off at high wave number faster than naive estimates based on the geometry of the defects, indicating non-trivial (anti-)correlations between the defects and the surrounding Goldstone boson field. We obtain source functions for Einstein-Boltzmann solvers from the UETCs, using a recently developed method that improves the 
modelling at the radiation-matter transition. We show that the interpolation function that mimics the transition is 
similar to other defect models, but not identical, 
confirming the non-universality of the interpolation function. 
The CMB power spectra for global strings and global monopoles have the same overall shape as 
those obtained using the non-linear $\sigma$-model approximation,
which is well captured by a large-$N$ calculation. 
However, the amplitudes are larger than the large-$N$ calculation would naively predict, and 
in the case of global strings much larger: a factor of 20 at the peak. 
Finally we compare the CMB power spectra with the latest CMB data in other to put limits on the allowed contribution to the temperature power spectrum at multipole $l = 10$ of $1.7$\% for global strings and $2.4$\% for global monopoles. 
These limits correspond to symmetry-breaking scales of $2.9\times 10^{15}\;\textrm{GeV}$ 
($6.3\times 10^{14}\;\textrm{GeV}$ with the expected logarithmic scaling of the effective string tension 
between the simulation time and decoupling)
and $6.4\times 10^{15}\;\textrm{GeV}$ respectively.
The bound on global strings is a significant one for the ultra-light axion scenario 
with axion masses $m_a \lesssim 10^{-28}$ eV.
These upper limits indicate that gravitational waves from global topological defects will not be observable at the 
gravitational wave observatory LISA.
}

\maketitle
\flushbottom
 

\section{Introduction}

Phase transitions in the early Universe may give rise to topological defects \cite{Vilenkin:2000jqa}, 
which are predicted in a wide range of high-energy physics models of the early Universe \cite{Vilenkin:2000jqa,Jeannerot:2003qv,Sarangi:2002yt,Jones:2003da}. 
Defects from symmetry breaking at the Grand Unified or inflation scales 
induce gravitational fluctuations of sufficient amplitude to give a  
characteristic signature in the cosmic microwave background (CMB).

The CMB power spectra of topological defects have been widely analysed. 
In some approaches, full-sky maps  \cite{Pen:1993nx,Allen:1997ag,Landriau:2002fx,Landriau:2003xf} are computed. 
Most use the scaling properties of defect networks to derive the power spectra from the  
Unequal Time Correlators (UETC) of the energy-momentum tensor of the classical field theory evolving in a cosmological background 
\cite{Pen:1997ae}.  
This method has been used to compute the CMB signatures of 
global defects in the non-linear sigma-model (NLSM) approximation \cite{Durrer:1998rw,Bevis:2004wk}, 
and cosmic strings in the Abelian Higgs model \cite{Bevis:2006mj,Urrestilla:2011gr,Bevis:2010gj}, 
and for semilocal strings \cite{Urrestilla:2007sf}.
Gauge cosmic string CMB power spectra have also been computed in the Nambu-Goto approximation \cite{Lazanu:2014eya,Lazanu:2014xxa} and by modelling strings as randomly moving string segments \cite{Albrecht:1997nt,Albrecht:1997mz,Battye:1997hu,Pogosian:1999np,Pogosian:2006hg,Pogosian:2008am}. 
Global strings have not been modelled in the Nambu-Goto approximation because their long-range interactions complicate the algorithm,  
although an economical way of incorporating the interactions has recently been put proposed \cite{Fleury:2016xrz}.

Using a similar approach gravitational wave power spectra have also been calculated \cite{JonesSmith:2007ne,Fenu:2013tea}. However, the authors of \cite{Figueroa:2012kw}, comparing predictions for GWs obtained from large-$N$ limit with the ones obtained from field theory simulations, showed that the NLSM prediction in the large-$N$ limit is significantly lower than the true value  
for models with $N<4$. 

In the last few years significant advances in the study of the CMB power spectrum using the UETC approach for Abelian Higgs strings were performed; using the biggest simulation boxes up to date and, among other improvements, studying the behaviour of the correlators across cosmological transitions \cite{Daverio:2015nva,Lizarraga:2016onn}. 

However, the CMB signature for O($2$) and O($3$) models have never been studied using the linear $\sigma$-model, 
which is required to capture the important contribution to the energy-momentum of the cores of the topological defects. 
Global string and monopole networks have been numerically analysed for other reasons: for example, in \cite{Moore:2001px,Yamaguchi:1999yp,Yamaguchi:1999dy,Hiramatsu:2010yu,Fleury:2015aca} the scaling properties of global string networks were studied. 
Axion strings are global strings, with axions as the (pseudo-)Goldstone bosons, and the radiation from global string networks has 
been examined in order to determine the dark matter axion density \cite{Hiramatsu:2010yu,Fleury:2015aca}. 

Similarly in \cite{Lopez-Eiguren:2016jsy} the network velocities of global monopoles were studied in detail. 
Therefore, the analysis of the CMB power spectrum generated by global strings and monopoles will be of great interest. As we have previously mentioned the NLSM predictions at large $N$ for GWs do not fit with O($2$) and O($3$) direct calculations. Therefore it is important to perform O($2$) and O($3$) field theory simulations to check whether they follow the CMB predictions from the O($N$) NLSM. 

In this work we perform field theory simulations of the O($2$) and O($3$) linear $\sigma$ models, 
measuring their scaling parameters to much greater accuracy than before, and determining the UETCs.
Armed with the UETCs, we calculate the CMB power spectra using the techniques detailed in \cite{Lizarraga:2016onn}. The power spectra are then used to compare with the predictions given by the NLSM and also with the spectrum coming from the analysis of the Abelian Higgs model. We also fit \Planck\ data with these predictions and obtain constraints on the models using a Monte Carlo analysis.

The paper is structured as follows: In section~\ref{sec:model} we present the model and we give an overview of the UETC approach. Then, we describe the procedure to obtain the UETCs from the simulations in section~\ref{sec:uetcs} and the computation of the source functions in section~\ref{sec:source}. Once we have the source functions we present the power spectra in section~\ref{sec:spectra}. Finally, in section~\ref{sec:fits} we show the fits and constraints and we conclude in section~\ref{sec:conclusions}.


\section{Model and Method overview}
\label{sec:model}
The simplest field theory model that contains global topological defects is the global O($N$) theory, where O($N$) global symmetry spontaneously breaks down to O($N-1$). A theory that gives rise to this kind of defects is the linear sigma model \cite{Goldstone:1961eq}, whose action is 
\be
\mathcal{S}=\int d^4 x \sqrt{-g} \Big( \frac{1}{2} \partial_{\mu} \Phi^i \partial^{\mu}\Phi^i- \frac{1}{4}\lambda(|\Phi|^2-\eta^2)^2 \Big),
\label{eq:ac}
\ee
where $\Phi^i$, $i=1,..,N$ are real fields, $|\Phi|\equiv \sqrt{\Phi^i \Phi^i}$ and $\lambda$ and $\eta$ are real constant parameters. In the symmetry breaking a massive particle with mass $m_s=\sqrt{2 \lambda}\eta$ arises as well as $N-1$ massless Goldstone bosons. 
The energy of local defects is  divergent with radius but in a cosmological situation this  is not catastrophic, since the energy divergence will be  cut-off  by neighbouring defects.

Since our aim is to study the dynamics of a network of global defects in an expanding universe, we consider a flat Friedmann-Robertson-Walker space-time with comoving coordinates:
\be
ds^2=a^2(\tau)(d\tau^2-dx^2-dy^2-dz^2),
\ee
where $a(\tau)$ is the cosmic scale factor and $\tau$ is conformal time. The equations of motion derived from (\ref{eq:ac}) are
\be
\ddot{\Phi}^i+2 \frac{\dot{a}}{a}\dot{\Phi}^i-\nabla^2 \Phi^i = -a^2 \lambda (\Phi^2-\eta^2)\Phi^i,
\label{eq:eom}
\ee
and the dots represent derivatives with respect to the conformal time $\tau$.

Since the size of the defect  is given by their inverse mass   $(\delta \sim m_s^{-1})$, a fixed length in physical units, which means that   in   comoving coordinates the size of the defects rapidly decreases. Thus, in order to have a longer dynamical range one has to use the Press-Ryden-Spergel method \cite{Press:1989yh}. 
This method makes the width of the defect controllable by turning the coupling constant into a time-dependent variable:
\be
\lambda=\lambda_0 a^{-2(1-s)},
\ee
where the parameter $s$ is the responsible to control the defect size. That is, when the $s=0$ the defect size is fixed in comoving coordinates and when $s=1$ we obtain the true case where the size of the defect is fixed in physical length.

This method and its extension for gauge theories has been widely checked 
\cite{Daverio:2015nva,Moore:2001px,Bevis:2010gj,Lopez-Eiguren:2016jsy}, where the errors due to its use are shown to be typically smaller than the statistical errors, or the systematic errors inherent to the discretization procedure.

The evolution of a defect network perturbs the background space-time; and those perturbations evolve and affect the contents of the universe, eventually creating CMB anisotropies. In contrast to inflationary perturbations, which were seeded primordially and then evolve ``passively'', defects induce perturbations actively during their whole existence. Those for Abelian Higgs cosmic strings are estimated to be roughly of the order of the magnitude of $G\mu$, where $G$ is Newton's constant and $\mu$ the string tension. Current bounds on $G\mu$ from CMB experiments constrain its value to be below $10^{-7}$ \cite{Ade:2013xla,Urrestilla:2011gr, Lizarraga:2014xza, Lazanu:2014xxa}. 

In order to describe the perturbations induced by defects energy-momentum correlations are appropriate statistical tools \cite{Turok:1996wa,Pen:1997ae,Durrer:2001cg,Bevis:2006mj}. Indeed, the two-point unequal time correlators of the energy-momentum tensor are the only objects needed to derive the power spectrum of CMB anisotropies. UETCs are defined as follows:
\be
U_{\lambda\kappa\mu\nu}(\mathbf{k},\tau,\tau')= \langle T_{\lambda\kappa}(\mathbf{k},\tau)T_{\mu\nu}(\mathbf{k},\tau')\rangle,
\label{eq:uetc}
\ee
where $T_{\alpha\beta}(\mathbf{k},\tau)$ is the energy momentum tensor.
 
The UETCs give the power spectra of cosmological perturbations when convolved with the appropriate Green's functions. In practice, they are decomposed into a set of functions derived from the eigenvectors of the UETCs, which are used as sources for an Einstein-Boltzmann integrator. The power spectrum of interest is reconstructed as the sum of power spectra from each of the source functions.

A considerable simplification occurs when the times $\tau$ and $\tau'$ are both in epochs during which the scale factor grows with the same constant power of conformal time. In this case the correlation functions do not depend on $k$, $\tau$ and $\tau'$ separately, but only on $k\tau$ and $k\tau'$. This behaviour is called scaling, and scaling correlators can be written as
\be
U_{ab}(\mathbf{k},\tau,\tau')=\frac{\eta^4}{\sqrt{\tau\tau'}}\frac{1}{V}\bar{C}_{ab}(k\sqrt{\tau\tau'},\tau'/\tau),
\label{scaUETC}
\ee
where now the indices $a$ and $b$ correspond to projections of the energy momentum tensor; specifically to its   independent components: two scalar (longitudinal gauge potential $\phi$ and $\psi$), one vector and one tensor. The overbar represents the scaling form of the UETC in a FLRW background. We will sometimes write $z=k \sqrt{\tau\tau'}$, $r=\tau'/\tau$. An alternative pair of scaling variables is $x, x'=k\tau,k\tau'$. A scaling UETC will have eigenvectors which depend on $k$ and $\tau$ only through the combination $x$.

Scaling is an essential feature for any kind of defect network to ensure their cosmological viability. Defect networks that exhibit scaling do not dominate the cosmological evolution over other species, but neither do they disappear. From the computational point of view it is also an immensely valuable property, since it allows to extrapolate the numerical observables obtained from limited reproductions of cosmological scenarios to required cosmological scales, which are well beyond current capabilities.

The UETCs are extracted from numerical simulations, where correlations between energy-momentum tensors at different stages of the evolution are calculated. After that, these functions, which are definite positive and symmetric, are diagonalized. The diagonalisation decomposes the UETCs into its eigenvalues and eigenfunctions,

\begin{equation}
\int_{\tau_{\rm i}}^{\tau_0} d\tau' C_{ab}(k,\tau,\tau') c_b^n(k,\tau') = \lambda_n(k)c_a^n(k,\tau).
\end{equation}
where $c_a^n$ are the eigenfunctions and $\lambda_n$ the eigenvalues associated to them.

Finally, in terms of these two ingredients, the source functions are defined in the following way:
\begin{equation}
s_a^n(k,\tau) =  \sqrt{\lambda_n(k)}c_a^n(k,\tau) \, .
\label{eq:source}
\end{equation}
These source functions are then what we will plug in to the Einstein-Boltzmann solvers to calculate the CMB anisotropy power spectra, for further details on the process see \cite{Durrer:2001cg,Bevis:2006mj,Daverio:2015nva}.

However, it should be noted that the scaling is broken near the cosmological transitions. That is, when the universe undergoes a transition from radiation-dominated era to matter-dominated era, or from matter-dominated era to the era dominated by dark energy the UETCs also depend explicitly on $\tau_{\rm eq}$ and $\tau_{\Lambda}$, the times of equal radiation and matter density, and equal matter and dark energy density.


\section{UETCs from the Simulations \label{sec:uetcs}}

In this section we present the details of the numerical simulations from which the scaling UETC data were collected. These scaling UETCs are the inputs for the next section, in which the eigenvector decomposition method is described.


\subsection{Simulation details}

In order to simulate the evolution of the global defects in a discrete box we discretise the action (\ref{eq:ac}), 
deriving discretised equations of motion  on 
a cartesian grid using a 3-point stencil for the spatial Laplacian and the leapfrog method for time evolution. 
The equations are evolved in $1024^3$ lattices with periodic boundary conditions using the LATfield2 library for parallel field theory simulations \cite{David:2015eya}. 
The periodic boundary conditions impose an upper limit on the time that the system can be evolved: 
beyond half a light-crossing time, when Goldstone bosons moving in opposite directions in the box will start to re-encounter each other.

Our simulation lattice has a comoving spatial separation of $dx=0.5$ and time steps of $dt=0.1$, in units where $\eta=1$. The simulation volume therefore has comoving size $L=512$. 
All simulations were run with $s=0$ and  $m_s(\tau) a(\tau) = 2\eta$, with $a = 1$ at the end of the simulation. 
We performed 5 individual runs in pure radiation and in pure matter dominated eras to determine the scaling form of the UETCs. We also performed runs across the radiation-matter cosmological transitions using the same parameters and initial conditions. 

We are interested in the scaling regime of the defect network, not on the details of the phase transition. Thus, the initial condition (at time $\tau_{\rm ini}$) used in the numerical simulation is not used to extract data; its only function is to drive the system to scaling in order to get as large as possible dynamical range.  We have found that for the present work a satisfactory  initial field configuration is given by  the scalar field velocities   set to zero and the scalar fields   set to be stationary Gaussian random fields with power spectrum
\be
P_{\phi}(\mathbf{k})=\frac{A}{1+(k\ell_{\phi})^2},
\ee
with $A$ chosen so that $\langle \Phi^2 \rangle=\eta^2$, and $\ell_{\phi}=5 \eta^{-1}$.

The UETCs cannot be calculated until  after the defects are formed and reach their scaling configuration. These early phases contain a huge amount of excess energy induced by the random initial conditions, therefore we smooth the field distribution by applying a period of diffusive evolution, with the second derivatives removed from the equations of motion and where the time step is $1/30$ in units where $\eta=1$. The length of  time  of the diffusive phase ($\tau_{\rm diff}$) varies, since the evolution of the equations of motion is different in each case. In all cases $\tau_{\rm diff}$    is determined in order to optimize the the dynamical range of the scaling evolution of the defect network.

After the diffusion period, the system relaxes into scaling, and we start to collect data from $\tau_{\rm ref}$ until the end of the simulation  $\tau_{\rm end}$.  We measure the UETC by recording the mean value of $C_{ab}(k,\tau_{\rm ref},\tau)$ for wavevectors binned in the range $2\pi(n-1)/L < |\mathbf{k}| \le 2n/L$ $(1\le n < N_{\rm b})$, with $N_{\rm b}=886$, and $n_{\rm out}$ number of outputs logarithmically-spaced times between $\tau_{\rm ref}$ and $\tau_{\rm end}$. The wavenumber of the $n$th bin $k_n$ is set to the mean value of $|\mathbf{k}|$ in that bin. Table  \ref{tab:sim-pro} shows the values of these parameters.

\begin{table*}
\begin{tabular}{|l | r | r|}
\hline
  & O($2$) & O($3$) \\
  \hline
  $\tau_{\rm ini}$ &  50 & 0 \\
  $\tau_{\rm diff}$ & 70 & 20 \\
  $\tau_{\rm ref}$ & 150 & 60 \\
  $\tau_{\rm end}$ & 300 & 250 \\
  $n_{\rm out}$ & 50 & 60 \\
  \hline
\end{tabular}
\caption{\label{tab:sim-pro}  The values of the time-related parameters, given in units where $\eta=1$. The simulations start at time $\tau_{\rm ini}$ and there is a period of diffusion until  $\tau_{\rm diff}$; the data are taken from   $\tau_{\rm ref}$  until $\tau_{\rm end}$ every $n_{\rm out}$ time-steps.}
\end{table*}

We also record the Equal Time Correlators (ETCs) at each time the UETC is evaluated, with which we can monitor the quality of scaling. Perfect scaling would mean that the ETCs collapse to a single line when plotted against $x=k\tau$.


\subsection{Scaling}
\label{s:Sca}

It is known that defining a length scale of the network is  convenient to track the state of the system and scaling. We will define two different length scales, one for each case of defect under study, \ie one for strings and another one for monopoles.

For the case of strings the comoving string separation $\xis$ has been identified as a useful quantity to determine compatible simulation stages \cite{Daverio:2015nva}. The string separation is defined in terms of the mean comoving string length $\ell_s$ in the comoving 
simulation volume $\mathcal{V}$ as
\be
\label{e:xisDef}
\xis=\sqrt{\frac{\mathcal{V}}{\ell_s}}.
\ee
The mean string length, $\ell_s$, is derived from estimators of the comoving length of string. 
One way of obtaining the length of strings is by summing the number of plaquettes pierced by 
strings. Such plaquettes are identified calculating the ``winding'' of the phase of the field around each plaquette of the lattice 
\cite{Vachaspati:1984dz}. 
We denote the string separation computed in this way as $\xiws$. 

An alternative way is to use local field theory estimators for the total string length \cite{Daverio:2015nva}. 
In our case we use the total comoving energy weighted by the potential $V$, 
\be
E_{V} = \mathcal{V} \frac{\int d^3 x T_{00} V}{\int d^3 x V} 
\ee
and the energy per unit length, $\musV$, also weighted with the potential,   
to define a string length estimator 
\be
\label{e:EVlen}
\Ls=\frac{E_{V}}{\musV}.
\ee
In order to obtain the potential weighted energy per unit length of global strings, we have solved numerically the static field equations for a straight string lying on the $z$-axis \cite{Vilenkin:2000jqa}. 
From the values of the profile functions we have calculated the weighted energy per unit length, which is $\musV=0.70\eta^2$.
 \begin{figure}[htbp]
    \centering
    \includegraphics[width=\textwidth]{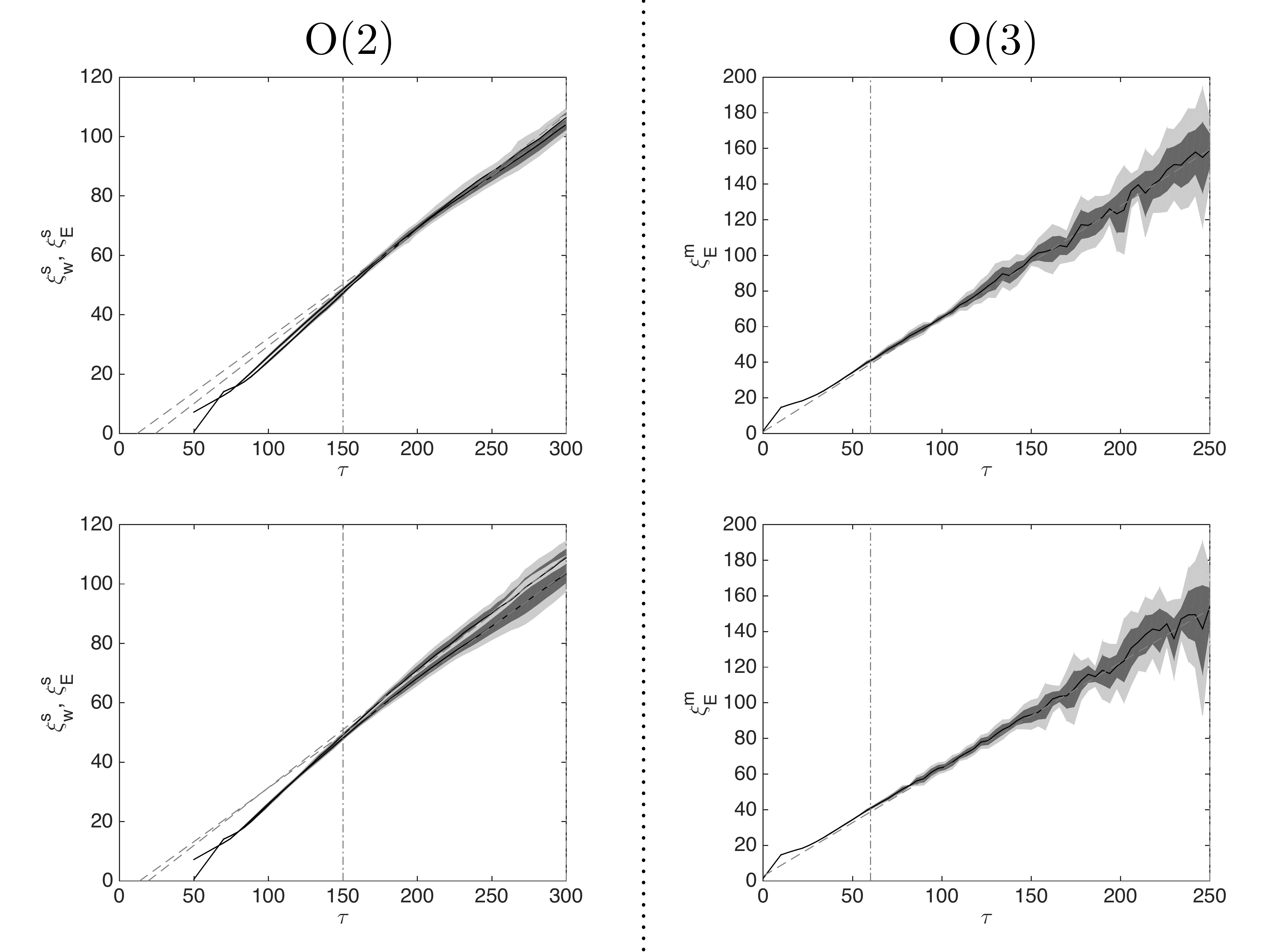} 
    \caption{In the left column: String separation $\xis$ (\ref{e:xisDef}) from simulations in radiation era (top figure) and matter era (bottom figure), with $\xiws$ obtained from the winding length measure and $\xiLs$ from the string length measure defined in (\ref{e:EVlen}). In the right column: Monopole separation $\xiLm$ (\ref{e:ximDef}) from simulations in radiation era (top figure) and matter era (bottom figure), obtained using 
    the monopole number estimate defined in (\ref{e:EVnum}).   
    }
 \label{fig:scaling}
 \end{figure}

Monopole networks can be characterized using the comoving monopole separation $\xim$. The monopole separation is defined in terms of the monopole number in the simulation volume $\mathcal{V}$ as
\be
\label{e:ximDef}
\xim=\Big( \frac{\mathcal{V}}{\mathcal{N}}\Big)^{1/3}
\ee 
where $\mathcal{N}$ is the total monopole number\footnote{Monopoles and antimonopoles contribute the same to 
the energy density, and are so equivalent in the energy-momentum tensor. Therefore, $\mathcal{N}$ is the sum of monopoles and antimonopoles. 
The net monopole charge in the simulation is exactly zero, due to the periodic boundary conditions.}.

The monopole number can be computed by directly obtaining the topological charge in each lattice 
cell of the simulation box \cite{Lopez-Eiguren:2016jsy,Antunes:2002ss}. 
It can also be obtained from a local field estimate 
\be
\label{e:EVnum}
\mathcal{N}=\frac{E_{V}}{\mumV},
\ee
where $\mumV$ 
is a energy of a monopole weighted with the potential $V$ and $E_{V}$ is the energy weighted with the potential. 
The weighted energy is computed in a similar way in which we have computed the weighted energy per unit length of global strings. That is, we have solved the equations of motion for a static monopole \cite{Vilenkin:2000jqa} using a relaxation method and then, using the profile functions we have calculated the weighted mass, which is $\mumV=2.33\eta$.

The computational cost of the field estimator is considerably lower, as the energy densities are being computed anyway, 
although it does slightly overestimate the monopole number
by about $10$\%.
This does not affect the linearity of $\beta^{\rm m}$ in the  
scaling regime, as we can see in Table~\ref{tab:betas3} where we have added the value of $\beta^{\rm m}_{\rm w}$ obtained from one simulation, and we restrict our separation measure to be the one from the field estimator.

As it was found in previous works, the asymptotic behaviour of the separations for both type of defects is very close to linear,
\be
\label{e:xiSca}
\xi \simeq \beta(\tau-\tau_{\rm off}),
\ee
where $\tau$ is the conformal time in the simulations and 
$\tau_{\rm off}$ is the time offset of the linear fit to the $\xi$ curve.\footnote{Without a time offset, plotting $\xi/\tau$ will produce  
an apparent time dependence in $\beta$, and one might wrongly conclude that the string network is not scaling.}

For the global string case the two different ways of computing $\xi^s$, $\xiws$ and $\xiLs$, give almost equal scaling behaviour and in the monopole case we have used $\xiLm$.
(see Fig. \ref{fig:scaling}).  
We have managed to find a combination of $\tau_{\rm ini}$ and $\ell_\phi$  
such that the time offset is almost zero in all realisations. 
We  define the mean slopes $\betaws$, $\betaLs$, and $\betaLm$
as the average of all different slopes from different realisations. 
Numerical values of the mean slopes obtained in the range $\tau \in [150 \; 300]$ for global strings can be found in Table \ref{tab:betas2} and in the range
$\tau \in [60 \; 250]$ for global monopoles can be found in Table \ref{tab:betas3}.

\begin{table*}
\begin{tabular}{|c|cccc|} 
\hline
& \multicolumn{4}{c|}{O($2$)} \\
\hline
-- & $\betaws$ &  $\betaLs$ & $\zeta$ & $\rho$ \\
 \hline
R & 0.36 $\pm$ 0.01 & 0.38 $\pm$ 0.01 &    1.7 $\pm$ 0.1   & 562 $\pm$ 9  \\

M & 0.36 $\pm$ 0.02 & 0.39 $\pm$ 0.02     &     0.72 $\pm$ 0.08  & 135 $\pm$ 9 \\
\hline
\end{tabular} 
  \caption{\label{tab:betas2} Numerical values of the different scaling parameters for global string networks 
  in the radiation (R) and matter (M) eras,
  obtained in the range $\tau \in [150 \; 300]$.
 The parameters $\beta$ are mean defect separations in units of the horizon length:  
 $\betaws$ is computed using the length of strings obtained from the number of windings, 
and $\betaLs$ is computed using the energy weighted with the potential, $E_{V}$.
The parameter $\zeta$ \cite{Moore:2001px} is a relative energy density and 
defined in \ref{eq:zeta}, while 
$\rho$ is a parameter defined in \cite{Pen:1993nx} for simulations in NLSM.
}
\end{table*}

\begin{table*}
\begin{tabular}{|c|cccc|} 
\hline
& \multicolumn{4}{c|}{O($3$)} \\
\hline
-- &  $\beta^{\rm m}_{\rm w}$       & $\betaLm$  & $\epsilon$ & $\rho$ \\
 \hline
R &  0.64 & 0.63 $\pm$ 0.03  &   1.26 $\pm$ 0.03 & 75 $\pm$ 2 \\

M &  0.59  & 0.60 $\pm$ 0.02   &  1.80 $\pm$ 0.04 & 28 $\pm$ 2\\
\hline
\end{tabular}
  \caption{\label{tab:betas3} Numerical values of the different scaling parameters for global monopole networks 
  in the radiation (R) and matter (M) eras,
  obtained in the range $\tau \in [60 \; 250]$.
 The parameters $\beta$ are mean defect separations in units of the horizon length:  
 $\betaws$ is computed using the number of monopoles obtained from the topological charge (data from only one simulation),  and  
$\betaLm$ is the mean separation of global monopoles, whose number density computed using 
the energy weighted with the potential, $E_{V}$.  
The parameter
$\epsilon$ \cite{Martins:2008zz} 
is the mean separation of monopole in units of the physical time $t$, while 
$\rho$ is a parameter defined in \cite{Pen:1993nx} for simulations in NLSM.
}
\end{table*}

We can translate the slopes of the mean separation to values of the parameter $\zeta$, 
proportional to the relative energy density, used in \cite{Moore:2001px}, 
defined as 
\be
\zeta=\frac{E_{V}}{\musV a^2(t)} t^2,
\label{eq:zeta}
\ee
where $t$ is the physical time. Our values of 
$\zeta$ can be seen in Table~\ref{tab:betas2} and Table~\ref{tab:betas3}. 
These values are compatible with the values given in \cite{Moore:2001px} in which 
$\zeta=2.0\pm 0.5$ for the radiation era, and the uncertainty is greatly reduced by the greater volume of our simulations.\footnote{Not allowing for a time offset (see Eq.~\ref{e:xiSca}) when extracting the scaling value of $\zeta$ can produce an apparent time dependence of $\zeta$, which is the reason for the apparent logarithmic growth observed in \cite{Fleury:2016xrz}.}

The values for the slope, $\betaLm$,
for the monopole case can also be seen in Table~\ref{tab:betas3}. These values are compatible with the values obtained in \cite{Lopez-Eiguren:2016jsy} where the authors found that $\beta_{\rm r}=0.72\pm 0.06$ in radiation era and $\beta_{\rm m}=0.65\pm0.04$ in the matter era.  In order to compare with the results obtained in \cite{Martins:2008zz} we define a scaling parameter $\epsilon$ by:
\be
\epsilon  = \frac{a(t)}{t}\Big( \frac{\mathcal{V}}{\mathcal{N}}\Big)^{1/3} 
\label{eq:epsilon}
\ee
where $t$ is the physical time. After the translation our values are shown in Table~\ref{tab:betas3}. These values are compatible with the values obtained in \cite{Martins:2008zz}, where two different sets of simulations were used. On the one hand, they used the simulations made in \cite{Yamaguchi:2001xn} and the results are $\epsilon_{\rm r}=1.3\pm0.4$ and $\epsilon_{\rm m}=1.6\pm0.1$. On the other hand, they have also used the simulations made in \cite{Bennett:1990xy} where the values for $\epsilon$ are $\epsilon_{\rm r}=1.3\pm0.2$ and $\epsilon_{\rm m}=1.9\pm0.2$.

We can also compare the scaling energy density with the values obtained in simulations of the O(2) and O(3) non-linear sigma model (NLSM) 
\cite{Pen:1993nx}, who defined a parameter
\be
\label{e:RhoDef}
\rho = \frac{E}{\mathcal{V}} \frac{\tau^2}{\eta^2},
\ee 
where $E = \int d^3 x T_{00}.$
We computed the scaling values of $\rho$ from the slopes of a linear fit of $\xi_E = \sqrt{\mu\mathcal{V}/E}$, and display the mean and standard deviations in Table~\ref{tab:betas2} and Table~\ref{tab:betas3}.
These are to be compared with the NLSM values of 
$68 \pm 7$  for O(2) and $24$ (no errors given) for O(3), both in matter era \cite{Pen:1993nx}.

Note that we have multiplied the scaling value $\rho = 14.5 \pm 1.5$ given in 
Ref.~\cite{Pen:1993nx} 
by $\ln(\xi_{\rm ref} m_s)$, in order to scale their results to our simulation volume.

The reason for this logarithm is that the global string energy per unit length 
should increase as $\mu\ln(\xi m_s)$ \cite{Vilenkin:2000jqa},
giving a logarithmic correction to scaling. In Ref.~\cite{Pen:1993nx} 
this logarithm is divided into the energy density in order to improve the scaling.
The value of the logarithm does not change significantly over the time range from which data is taken in our simulations, 
so we can extract scaling correlators without this compensation.  

The difference in scaling densities for string indicates that NLSM simulations are missing an important energy contribution from the string cores.
Monopole cores, on the other hand, make little difference to the energy density, as their contribution decreases 
as $\tau^{-3}$, and so one expects the agreement between the NLSM and the linear $\sigma$-model to be good.

\subsection{Energy momentum correlators}
\label{subsec:uetcs}

\begin{figure}[htbp]
    \centering
    \includegraphics[width=0.49\textwidth]{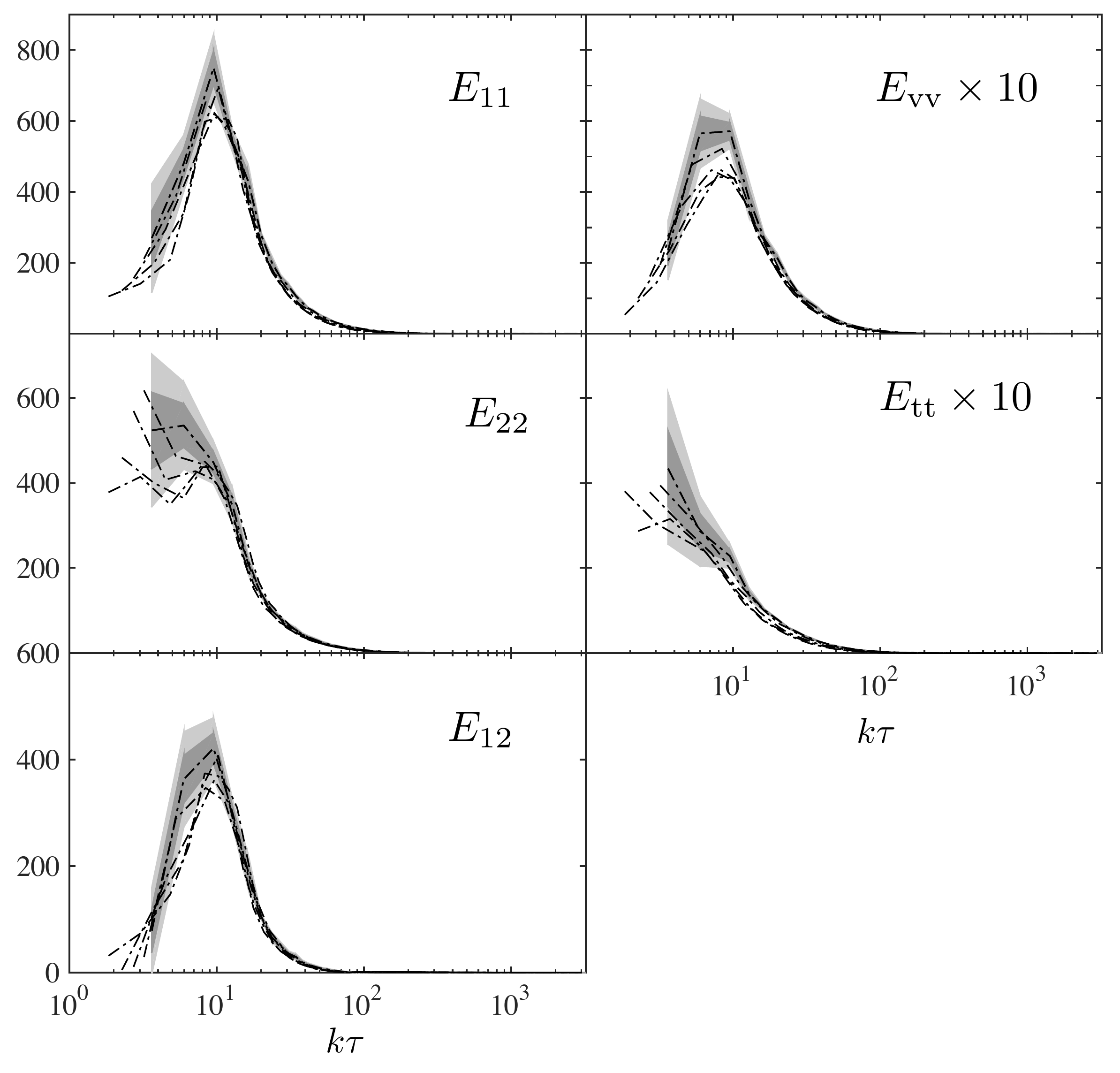} 
    \includegraphics[width=0.49\textwidth]{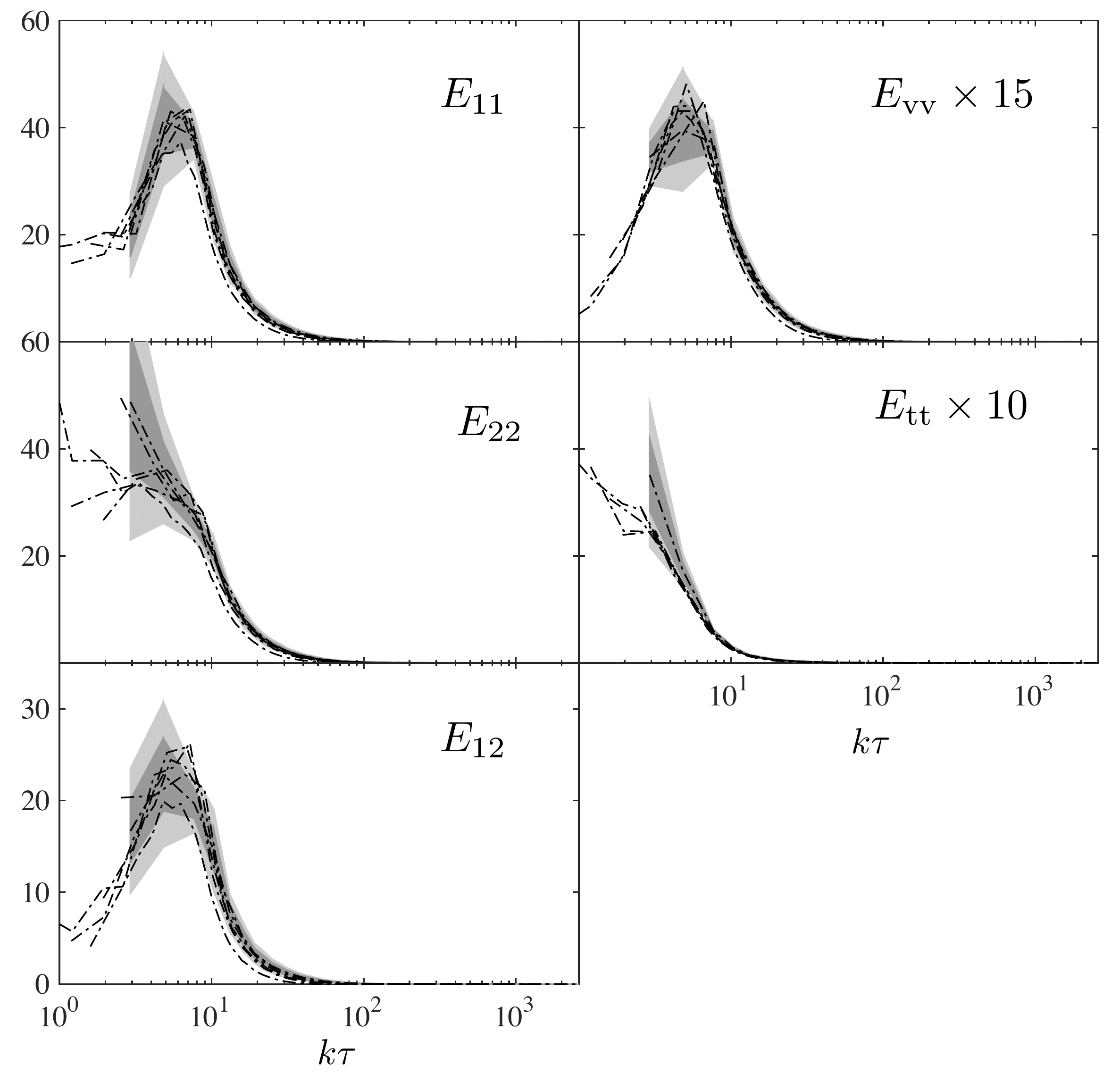} 
     \caption{ETCs for global strings (left pane) and for global monopoles (right pane) in radiation era. Shaded regions correspond to $1\sigma$ and $2\sigma$  deviations calculated at $\tau_{\rm end}$. The 5 dashed lines in each case correspond to ETCs at intermediate times, in units of $\eta^{-1}$: 150, 185, 222, 261 and 296 for O($2$) strings and 60, 97, 130, 158 and 205 for O($3$) monopoles.
     }
    \label{fig:etcs}
 \end{figure}

The ETCs of the energy momentum tensor give a more detailed test of scaling. In Fig.~\ref{fig:etcs} we show the ETCs for global strings and global monopoles in radiation era for the whole period of time recorded. We show the ETCs at $\tau_{\rm end}$ with shaded regions that represent $1\sigma$ and $2\sigma$ levels obtained averaging over 5 realizations  and the dashed lines correspond to intermediate times: 150, 185, 222, 261 and 296 for O($2$) and 60, 97, 130, 158 and 205 for O($3$) (in units of $\eta^{-1}$). The behaviour in the matter era is similar to the behaviour in the radiation era. The figures show that at small scales the ETCs collapse to a single line, though this behaviour is not so clear at low-k$\tau$'s.

\begin{figure}[htbp]
    \centering
    \includegraphics[width=0.49\textwidth]{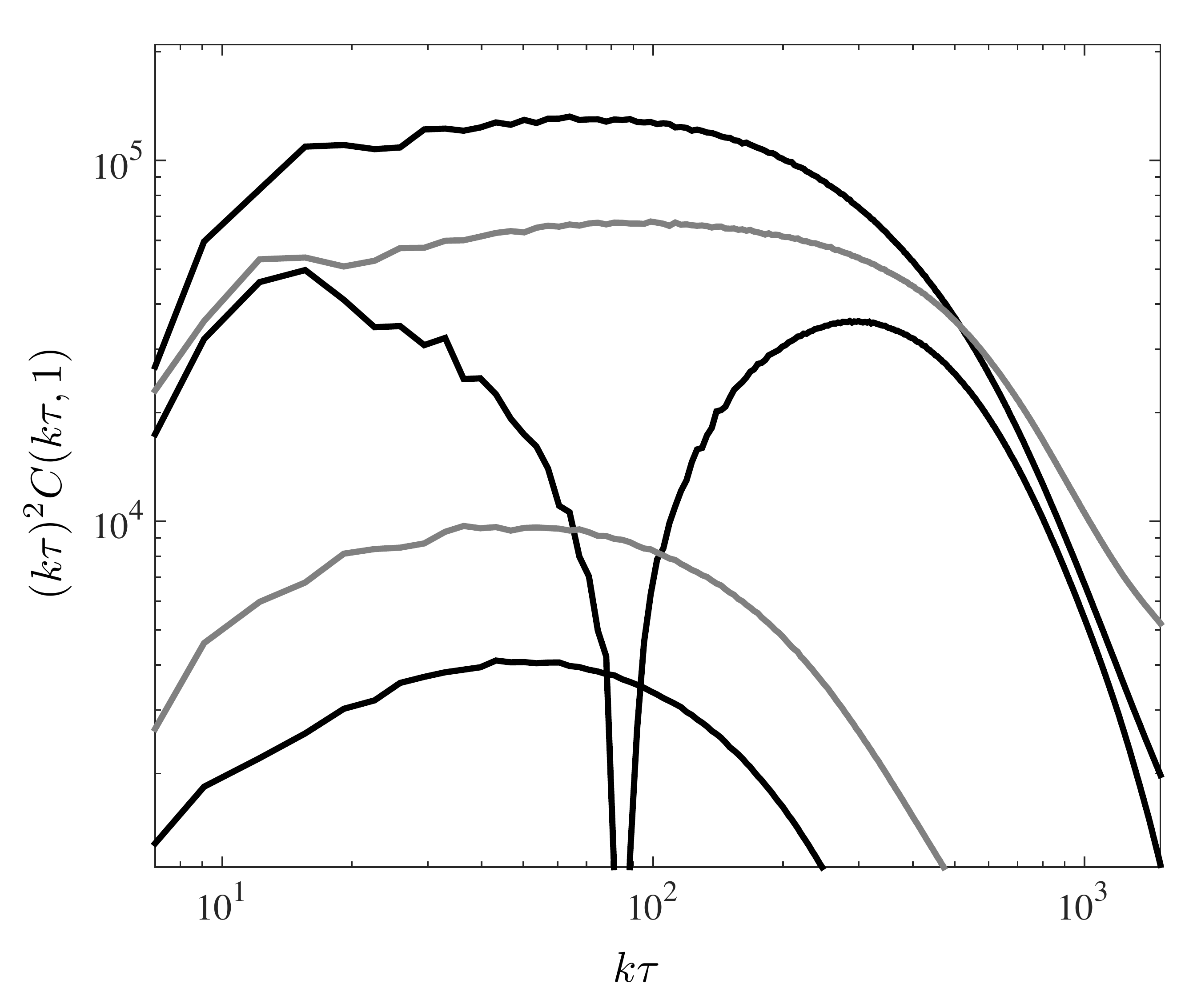}
    \includegraphics[width=0.49\textwidth]{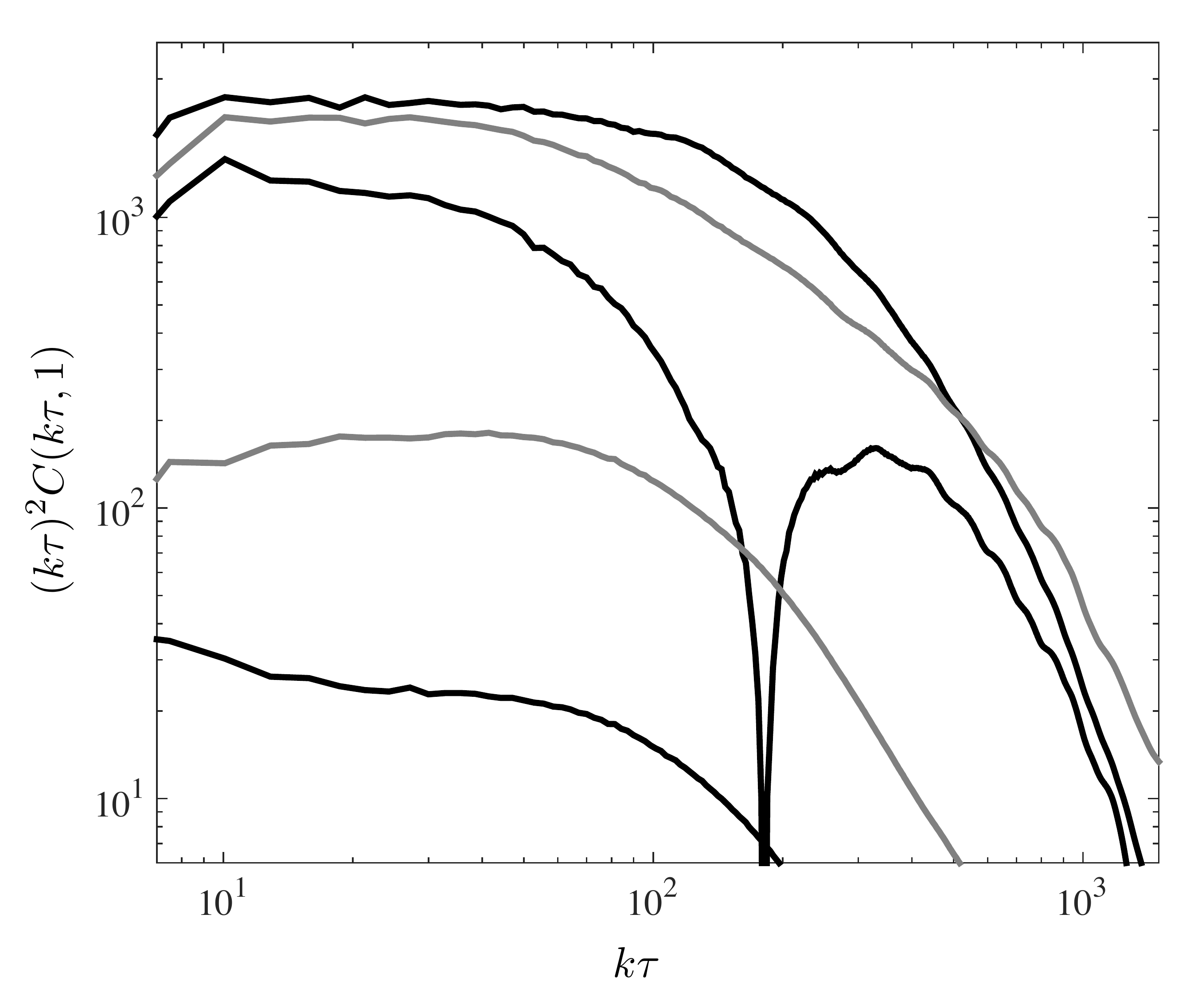} 
     \caption{Logarithmic plots for all five equal time correlators at the end of radiation era simulations for global strings  $\tau\simeq300$ (left pane) and global monopoles $\tau\simeq250$ (right pane). Note that the correlators are multiplied by $(k\tau)^2$ to demonstrate that the fall-of beyond the peak is      approximately $k^{-2}$.       The colour scheme in both cases is the following (from top to bottom): $C_{11}$ black, $C_{22}$ gray, $C_{12}$ black, $C_{\rm vv}$ black and $C_{\rm tt}$ gray.}
    \label{fig:etcsPL}
 \end{figure}

We have also studied the decay of the ETCs at small scales,  and figure~\ref{fig:etcsPL} shows the ETCs multiplied by $(k\tau)^2$ in logarithmic scale. We observe that for global strings as well as for global monopoles the ETCs decay roughly as $k^{-2}$, which contrasts with the expected string-like $k^{-1}$ \cite{Vincent:1996qr} and point-like $k^0$ behaviour. Instead, it is consistent with the $k^{-2}$ short-distance power spectrum induced by randomly-placed disks. We have looked at various visualisations of the energy-momentum tensor, and we have been unable to find one which gave a good impression of the sheet-like structures suggested by the power spectrum. It would interesting to explore the implied non-trivial correlations between defects and the surrounding Goldstone boson cloud.

\begin{figure}[htbp]
   \centering
    \includegraphics[width=6.5cm]{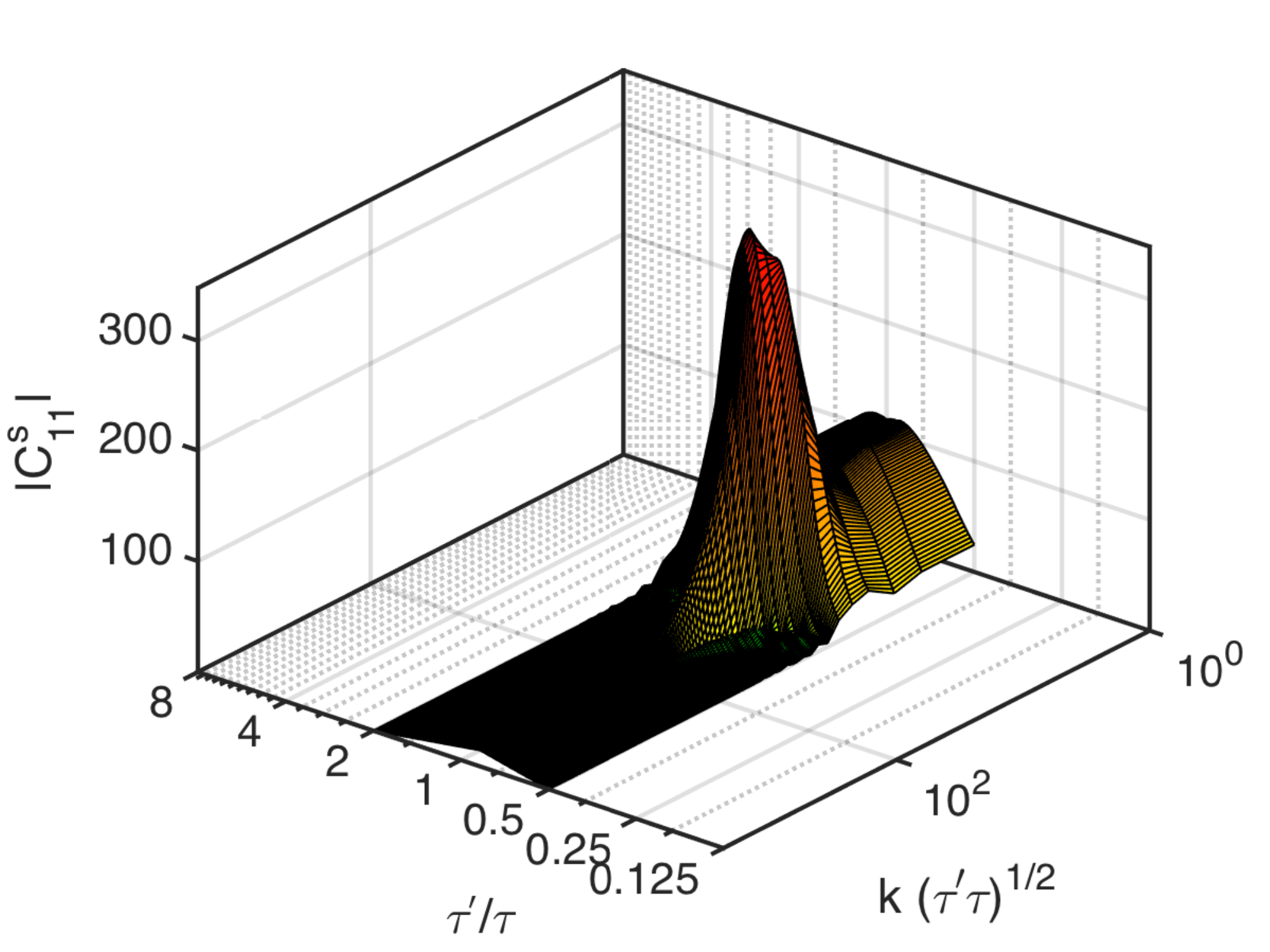} 
   \includegraphics[width=6.5cm]{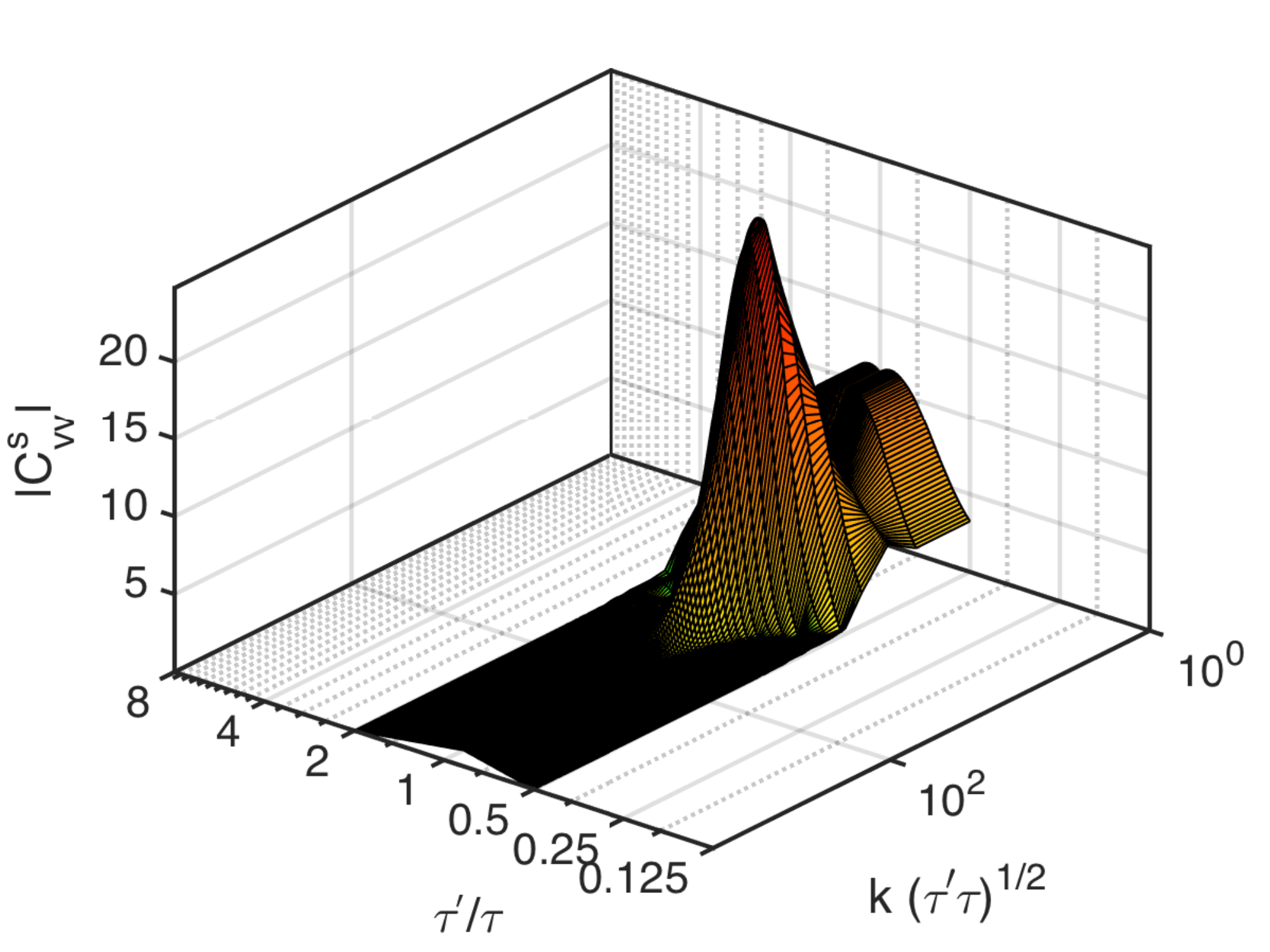} 
   \includegraphics[width=6.5cm]{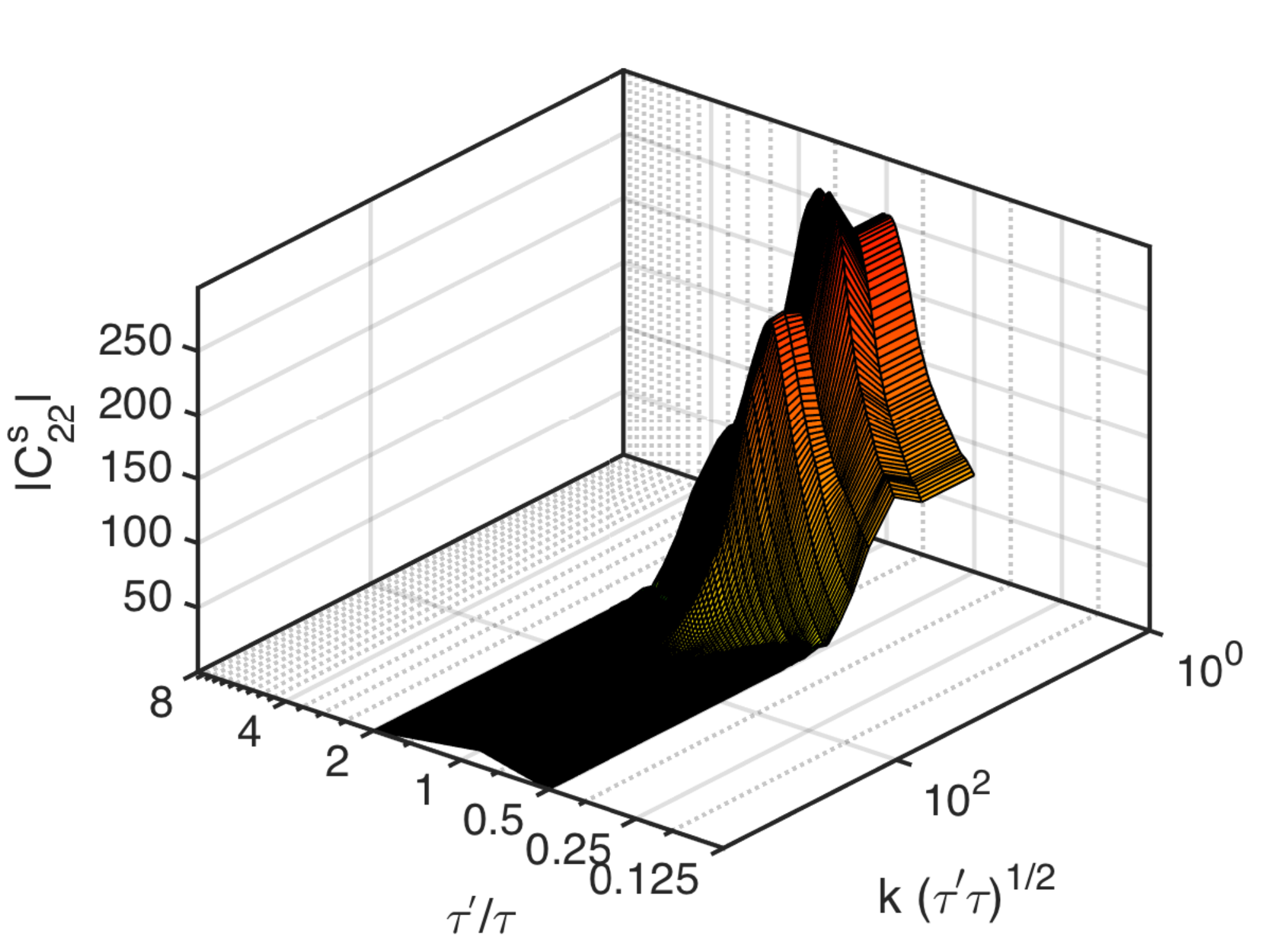} 
   \includegraphics[width=6.5cm]{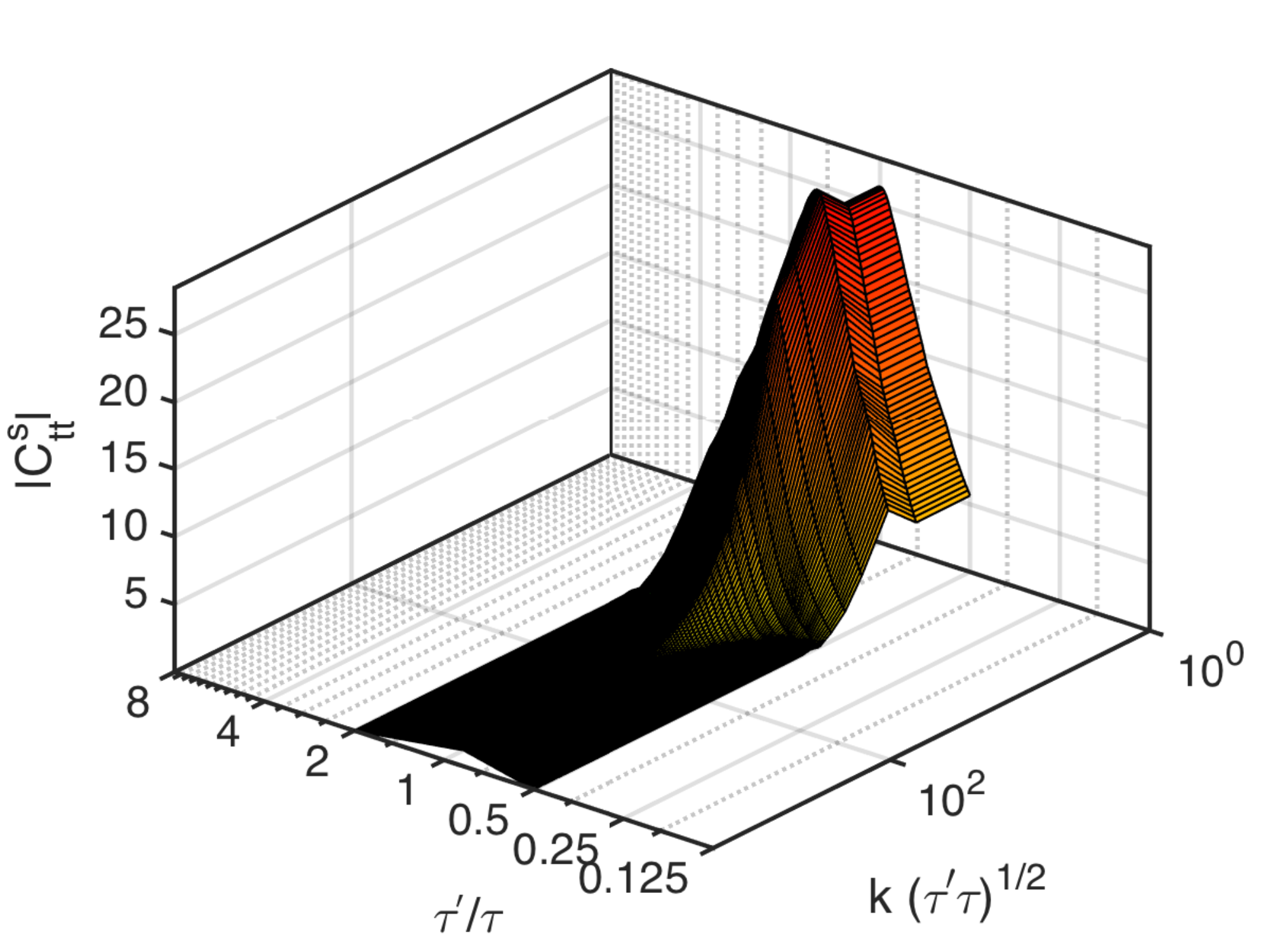} 
   \includegraphics[width=6.5cm]{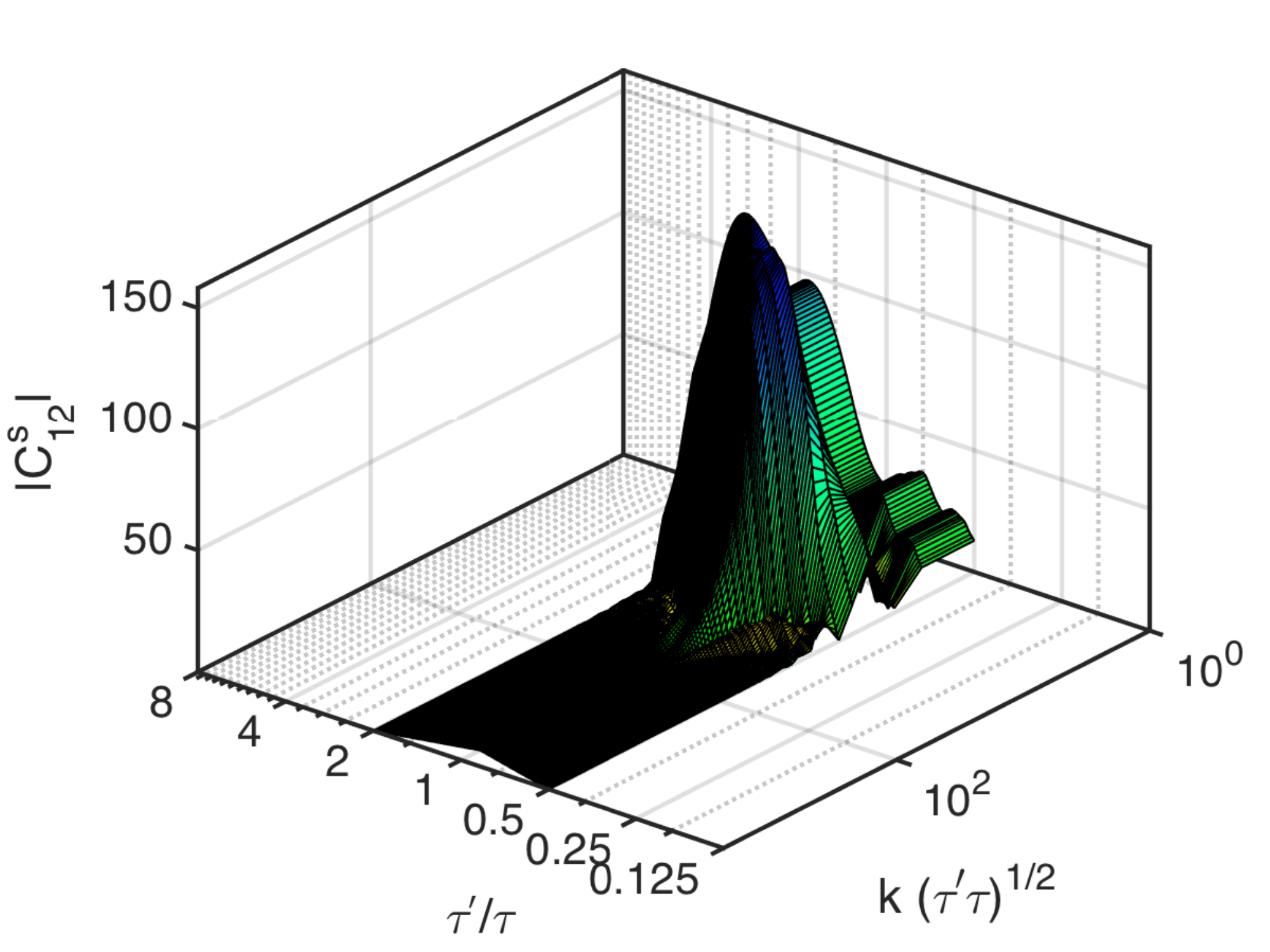}
    \hbox to 60mm{}
    \caption{Full set of scaling O($2$) UETCs for the matter era, calculated  averaging over 5 runs. See Section~\ref{sec:model} for definition of UETCs.} 
    \label{fig:uetc-nf2-rad2}
 \end{figure}

\begin{figure}[htbp]
    \centering
    \includegraphics[width=6.5cm]{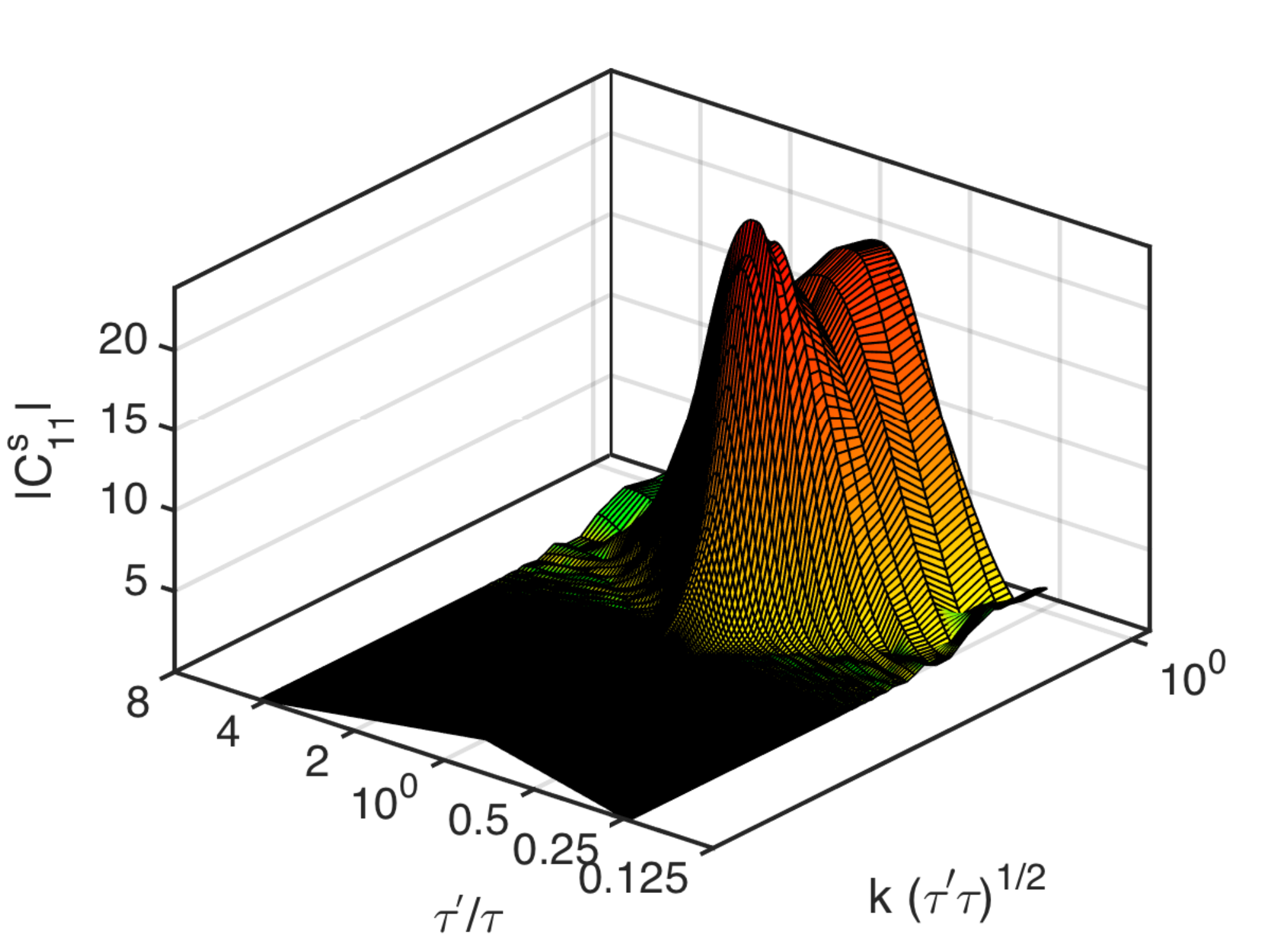} 
   \includegraphics[width=6.5cm]{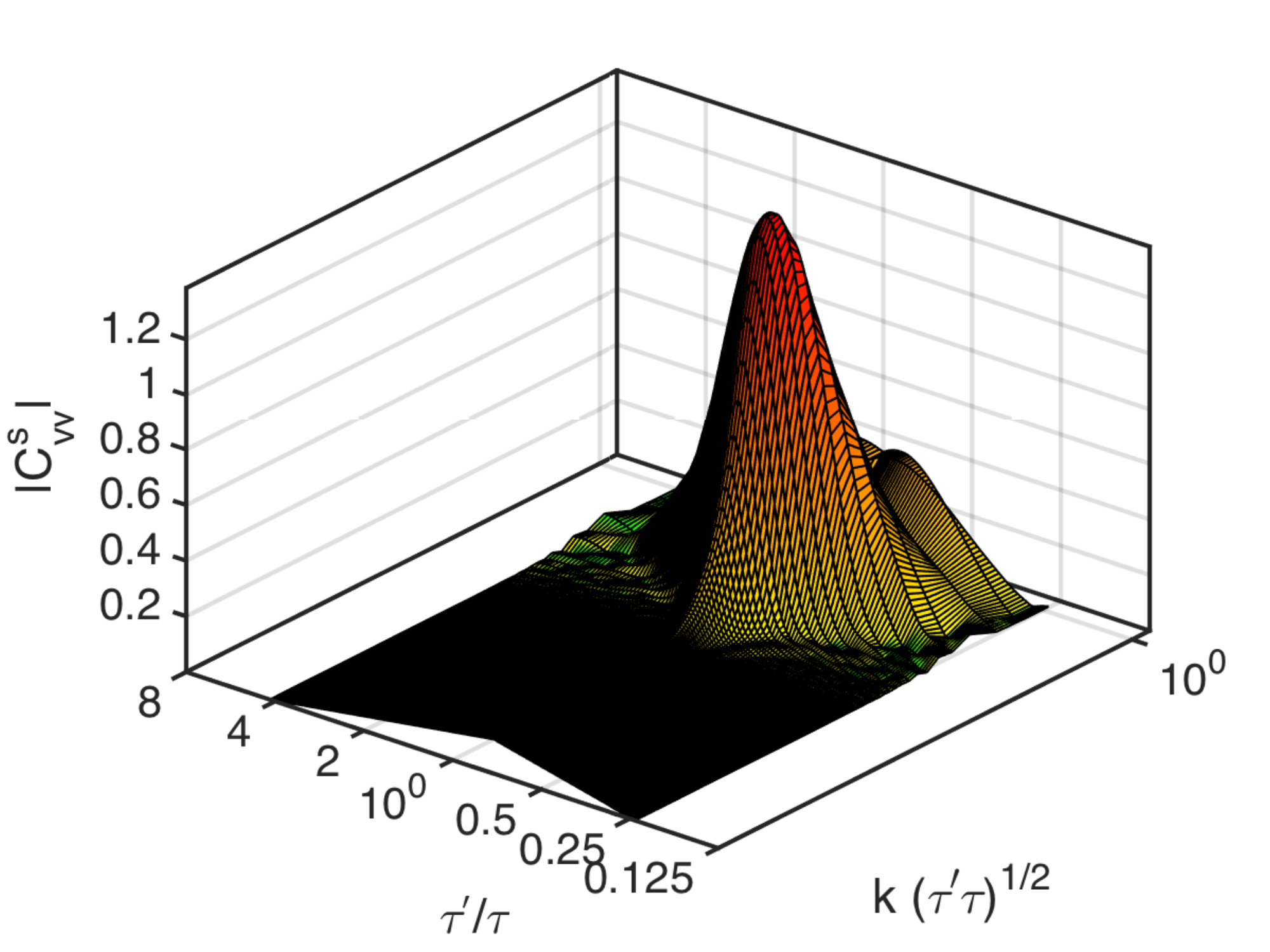} 
   \includegraphics[width=6.5cm]{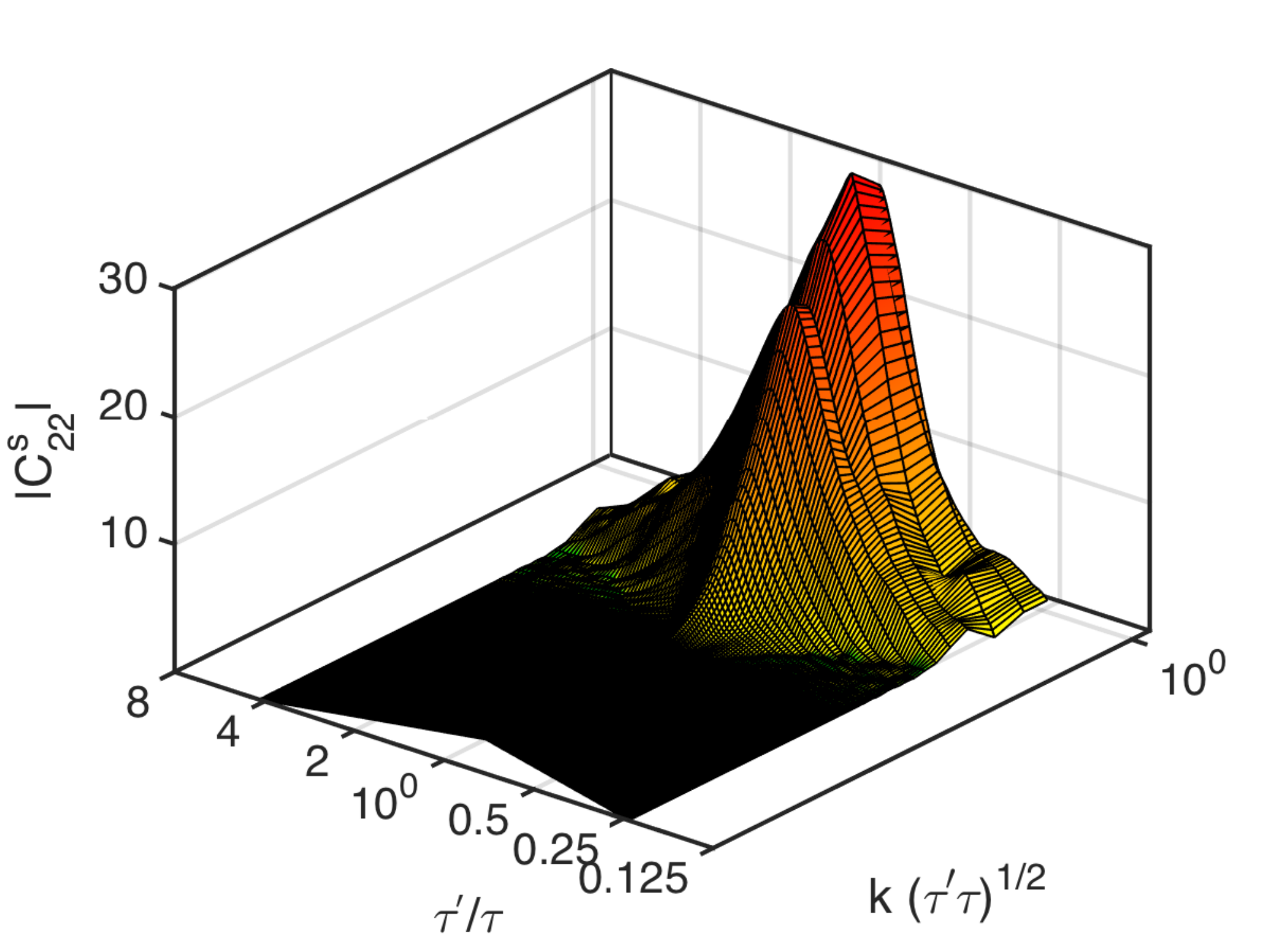}
   \includegraphics[width=6.5cm]{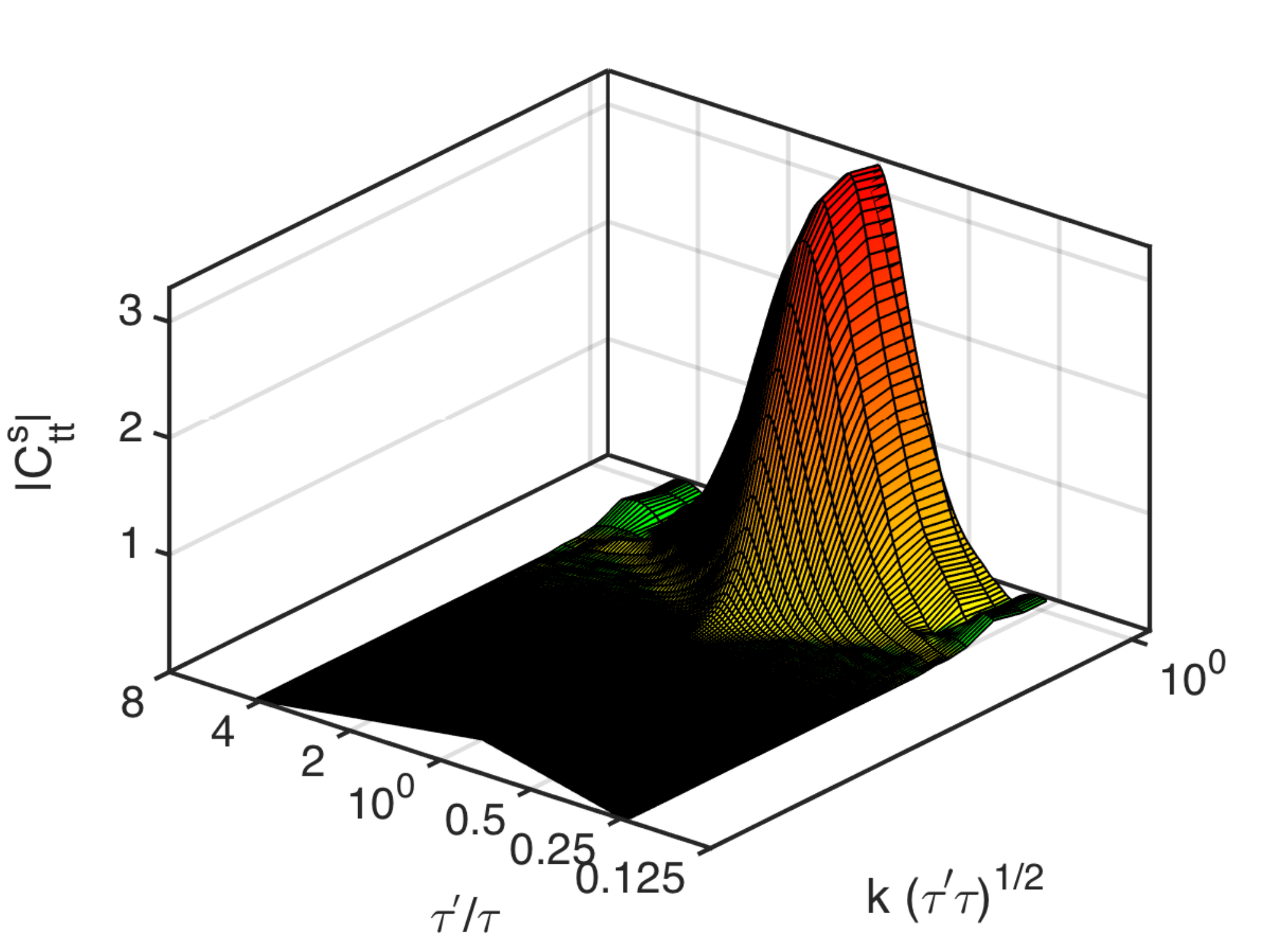} 
   \includegraphics[width=6.5cm]{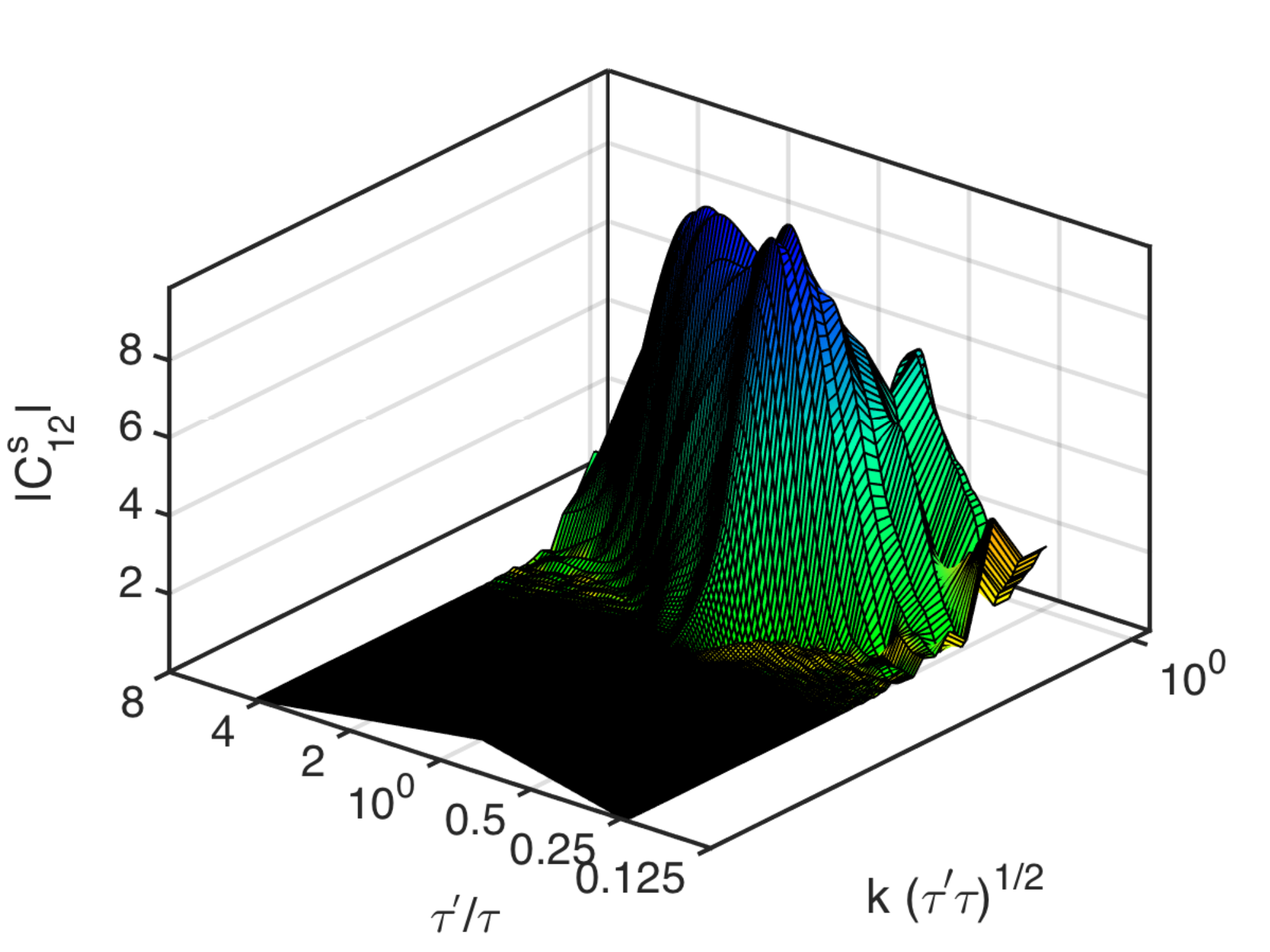}
    \hbox to 60mm{}
    \caption{Full set of scaling O($3$) UETCs for the matter era, calculated averaging over  5 runs. See Section~\ref{sec:model} for definition of UETCs.} 
    \label{fig:uetc-nf2-rad3}
 \end{figure}

Since the offset is consistent with zero in our simulations,  
it is straightforward to average the UETCs obtained from different realizations.
Figure~\ref{fig:uetc-nf2-rad2} shows the averaged Matter UETCs for global strings, and Fig.~\ref{fig:uetc-nf2-rad3} shows the corresponding one for global monopoles.

Figures \ref{fig:uetc-nf2-rad2} and \ref{fig:uetc-nf2-rad3} show that the amplitudes of the correlators (in matter era) of O($2$) strings are much bigger than the amplitude of O($3$) UETCs. Note that the UETCs for radiation era, in both defect types under analysis, have the same shape but the power is bigger than in the matter era. Similarly, if we compare global string correlators with the UETCs obtained from simulations of the Abelian Higgs model presented in \cite{Daverio:2015nva}, we observe that the general shape is similar while the amplitude is slightly higher for the global case. In both cases we use units where the vacuum expectation value of the scalar field is 1. Note that the normalisation convention is different for complex and real scalar fields.

 
\section{Computation of source functions \label{sec:source} } 

It has been established in the previous section that global strings and monopoles evolve in the scaling regime for most of the time reproduced by our simulations. As it is known, and also mentioned in section~\ref{sec:model}, the scaling can be used to extrapolate results derived from numerical simulations of different type of defects to the required cosmological scales.  The universe undergoes two transitions during times of interest, say the transition from radiation-dominated era to matter-dominated era and the transition from matter-domination to $\Lambda$-domination. In this work we will not consider the transition from matter-domination to $\Lambda$-domination, since its effect is rather small, as shown in \cite{Lizarraga:2016onn}. Therefore, perfect scaling is not a feature of networks evolving in our universe, this is why the scales imposed by the transitions must be also considered.

UETCs  must also depend on the scales imposed by the transitions. This means that in general the correlators will depend explicitly on $\tau_{\rm eq}$, in other words, the true (non-scaling) UETCs are functions of three different dimensionless variables, 
which can be chosen to be $k\tau$, $k\tau'$ and $\sqrt{\tau\tau'}/\tau_{\rm eq}$. One has to determine a method which captures the information of the transitions and include it in the computation of the source functions. There are several proposals in the literature for  performing this estimation \cite{Fenu:2013tea, Lizarraga:2016onn, Bevis:2010gj,Daverio:2015nva}, all of which were compared in  \cite{Daverio:2015nva}.

In this work we will follow the fixed-$k$ interpolation method proposed in \cite{Daverio:2015nva}: the UETCs are thought of as symmetric functions of $\tau$ and $\tau'$ for a given $k$. This approach has several advantages: it preserves orthogonality of the source functions during defects' whole existence and reproduces better the UETCs at cosmological transitions. Moreover, it also fits very well into the scheme used by Einstein-Boltzmann codes, which solve the perturbation equations with an outer loop over $k$ and an inner time integration for fixed values of $k$. For further details we refer the reader to \cite{Daverio:2015nva}.

The true UETCs $C_{ab}(k,\tau,\tau')$ are constructed from the mixture of the scaling matter and radiation correlators, extracted from our simulations, at each value of $k$. The relative mixture of matter and radiation UETCs is determined by $\tau/\tau_{\rm eq}$ and $\tau'/\tau_{\rm eq}$. An explicitly symmetric proposal for the UETCs which models this behaviour across the radiation-matter transition is the following \cite{Daverio:2015nva}:
\be
C_{ab}(k\tau,k\tau',\sqrt{\tau\tau'}/\tau_{\rm eq})= f\left( \frac{\sqrt{\tau\tau'}}{\tau_{\rm eq}}\right) \bar{C}_{ab}^{\rm R}(k\tau,k\tau')+\left( 1-f\left( \frac{\sqrt{\tau\tau'}}{\tau_{\rm eq}}\right) \right) \bar{C}_{ab}^{\rm M}(k\tau,k\tau').
\label{eq:trans}
\ee
It approximates the UETC in the entire region by the linear combination of pure radiation and pure matter era scaling correlators balancing the contribution of each by an interpolating function $f$. At extreme values of $\tau / \tau_{\rm eq}$ we recover functions that correspond to matter ($\tau/\tau_{\rm eq}\gg 1$) and radiation ($\tau/\tau_{\rm eq}\ll1$) dominations

We note that the source functions for the EB integrators at a given $k$ are now just the eigenvectors of these model UETCs, multiplied by the square root of the associated eigenvalues, and so they are indeed orthogonal, see Eq.~(\ref{eq:source}).

In order to establish the form of the interpolating function, we perform numerical simulations of O($2$) and O($3$) defects at cosmological transitions. The interpolating function can be defined in the following way in terms of the equal-time correlators (ETC) $E_{ab}(k,\tau)=C_{ab}(k,\tau,\tau)$  \cite{Fenu:2013tea}:
\be
f_{ab}(k,\tau)=\frac{E_{ab}^{\rm RM}(k,\tau)-\bar{E}_{ab}^{\rm M}(k\tau)}{\bar{E}_{ab}^{\rm R}(k\tau)-\bar{E}_{ab}^{\rm M}(k\tau)} \quad \forall k,
\label{eq:trans-fun}
\ee
where $\bar{E}^{\rm R}(k\tau)$ and $\bar{E}^{\rm M}(k\tau)$ are the scaling ETCs in the radiation and matter eras respectively, and $E^{\rm RM}(k,\tau)$ is the true ETC measured from the simulations performed during the transition.

\begin{figure}[htbp]
    \centering
    \includegraphics[width=15cm]{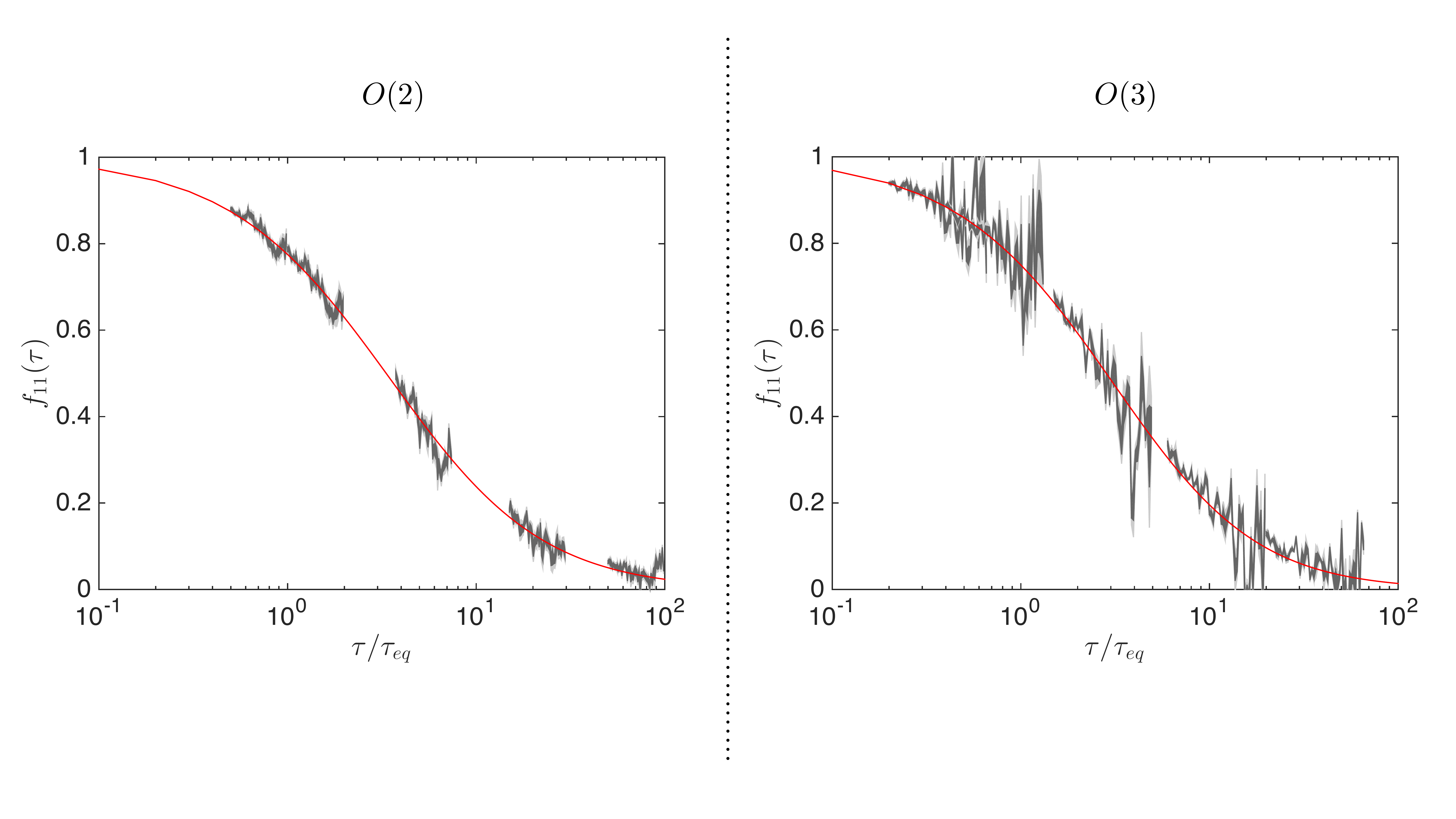} 
       \caption{UETC interpolation functions derived from simulations performed during the radiation-matter transition corresponding to global strings (left panel) and global monopoles (right panel)(thick grey line). The five patches correspond to simulations with $\tau_{\rm eq}=3,\ 10,\ 40,\ 150$ and $300$. The shaded regions represent the $1\sigma$ and $2\sigma$ deviations from the mean value of the function obtained from Eq.~(\ref{eq:trans-fun}) calculated from averaging over $k$, while the red line corresponds to the best-fit given by the function expressed in Eq.~(\ref{eq:ftau}). In both panels the correlator used is $E_{11}$. 
    \label{fig:trans-fun}}
 \end{figure}

We extracted ETCs from the simulations with $\tau_{\rm eq}= 3,\ 10,\ 40,\ 150$ and $300$ (see Table~\ref{tab-runs}), and used Eq.~(\ref{eq:trans-fun}) to compute the function. Fig. \ref{fig:trans-fun} shows the results obtained for $E_{11}$ correlators for global strings (left panel) and global monopoles (right panel), the transition functions for the rest of the correlators are similar to those shown in the figure. The five grey shaded regions represent the raw transition functions obtained during the five transition periods simulated. The two grey levels indicate $1\sigma$ and $2\sigma$ deviations from the mean value calculated averaging over a set of wavevectors: $1.5 < |\mathbf{k}| < 3.5$ and $3 < |\mathbf{k}| < 5$ respectively.

\begin{table*}
\renewcommand{\arraystretch}{1.2}
\begin{tabular}{|c|c|c|c|c|c| }
\hline
$\tau_{\rm eq}$ & 300 & 150 & 40 & 10 & 3 \\
$\tau_{ \rm ref}/\tau_{\rm eq}$ & 0.5 & 1.0 & 3.75 & 15 & 50 \\
$\tau_{\rm end}/\tau_{\rm eq}$ & 1.00 & 2.0 & 7.5 & 50 & 100 \\
$\alpha(\tau_{\rm ref})$ & 1.09 & 1.17 & 1.44 & 1.76 & 1.91\\
$\alpha(\tau_{\rm end})$ & 1.17 & 1.29 & 1.60 & 1.86 & 1.95 \\ 
\hline
\end{tabular}
\caption{\label{tab-runs} Selected parameters for simulations across the radiation-matter transition. The parameters are the conformal time of matter-radiation equality, $\tau_{\rm eq}$, in units of $\eta^{-1}$, the ratio of the reference time $\tau_{\rm ref}$ for UETC data taking and the simulation end time $\tau_{\rm end}$ to $\tau_{\rm eq}$, and the expansion rate $\alpha=d \ln a/ d \ln \tau$ at $\tau_{\rm ref}$ and $\tau_{\rm end}$.}
\end{table*}

The interpolating functions derived from our simulations confirm what previous analysis of the behaviour of the energy-momentum correlators at cosmological transitions showed: scale independence of the interpolating function. The deviations from the mean value represented by two grey levels, though they are somewhat bigger for monopoles, are not significant in either case. This demonstrates that the interpolating functions depend only on time to a good approximation. The rest of the correlators (not shown) support the scale independence statement, with the same function, which we shall write $f(\tau)$.

We fit the data using the same form used in \cite{Fenu:2013tea,Daverio:2015nva}, which is 
\be
f(\tau)= \left( 1+ \gamma \frac{\tau}{\tau_{\rm eq}} \right)^{\kappa}, 
\label{eq:ftau}
\ee
where $\gamma$ and $\kappa$ are the parameters to be determined by the fitting process.

Table~\ref{tab:trans} shows the mean values and standard deviations for the parameters of Eq.~\ref{eq:ftau}. The mean and standard deviations are obtained averaging over different realizations and over different correlators, since it has been observed that in a good approximation the interpolating function is the same for all correlators. The best-fit obtained fitting data is also included in Fig.~\ref{fig:trans-fun}.

\begin{table*}
\renewcommand{\arraystretch}{1.2}
\begin{tabular}{|c|c|c|}
 \hline
 &$\gamma$ & $\kappa$ \\
 \hline
  O($2$) & 0.26 $\pm$ 0.03 & -1.15 $\pm$ 0.02 \\
  O($3$) & 0.23 $\pm$ 0.05 &  -1.4  $ \pm$ 0.2 \\
  \hline
  \end{tabular}
  \caption{\label{tab:trans}  Mean values together with the standard deviations for parameters $\gamma$ and $\kappa$ of Eq.~(\ref{eq:ftau}) needed to reproduce the radiation-matter transition.}
\end{table*}

In \cite{Fenu:2013tea} it was proposed that the interpolation function should be universal in all defect models, with $\gamma = 0.25$ and $\kappa = -2$. In \cite{Daverio:2015nva} it was found that for the Abelian Higgs model, 
the interpolation function had the same form, but with $\gamma = 0.25$ and $\kappa = -1$, which is already a counterexample.
From our results we see that the interpolation function is not universal even within O($N$) defect models.
However for bigger $N$'s the value of $\kappa$ is bigger, which might be a sign of a trend. It would be interesting to test whether increasing the value of $N$ we eventually get the value proposed in \cite{Fenu:2013tea}.

Finally, having determined how the transitions has to be performed for the two defects analyzed in this paper, we diagonalise the true non-scaling UETCs Eq.~(\ref{eq:trans}) and obtain the source functions Eq.~(\ref{eq:source}) that will be used for the CMB power spectra calculation, as we describe in the next section.


\section{Power Spectra \label{sec:spectra}}

In the previous section we have defined the source functions for the global strings and monopoles. Inserting these functions into a source enabled Einstein-Boltzmann (EB) solver we can compute the contributions to CMB power spectra due to the presence of global defects. In our case the EB solver we have used is the source enabled version of CMBEASY \cite{Doran:2003sy}, \ie the code has been additionally modified to handle source functions of that we have explained in the previous section.

The cosmological parameters used for these calculations are the best-fit values obtained by the Planck collaboration \cite{Ade:2015xua}: $h=0.6726$, $\Omega_{\rm b}h^2=0.02225$, $\Omega_{\Lambda}=0.6844$ and reionization optical depth $\tau_{\rm re}=0.079$. After diagonalisation, the total contribution of defects under analysis to temperature and polarization anisotropies is calculated summing the contribution of each individual source functions, where 130 source functions are summed in each case. 

\begin{figure}[htbp]
    \centering
    \includegraphics[width=15cm]{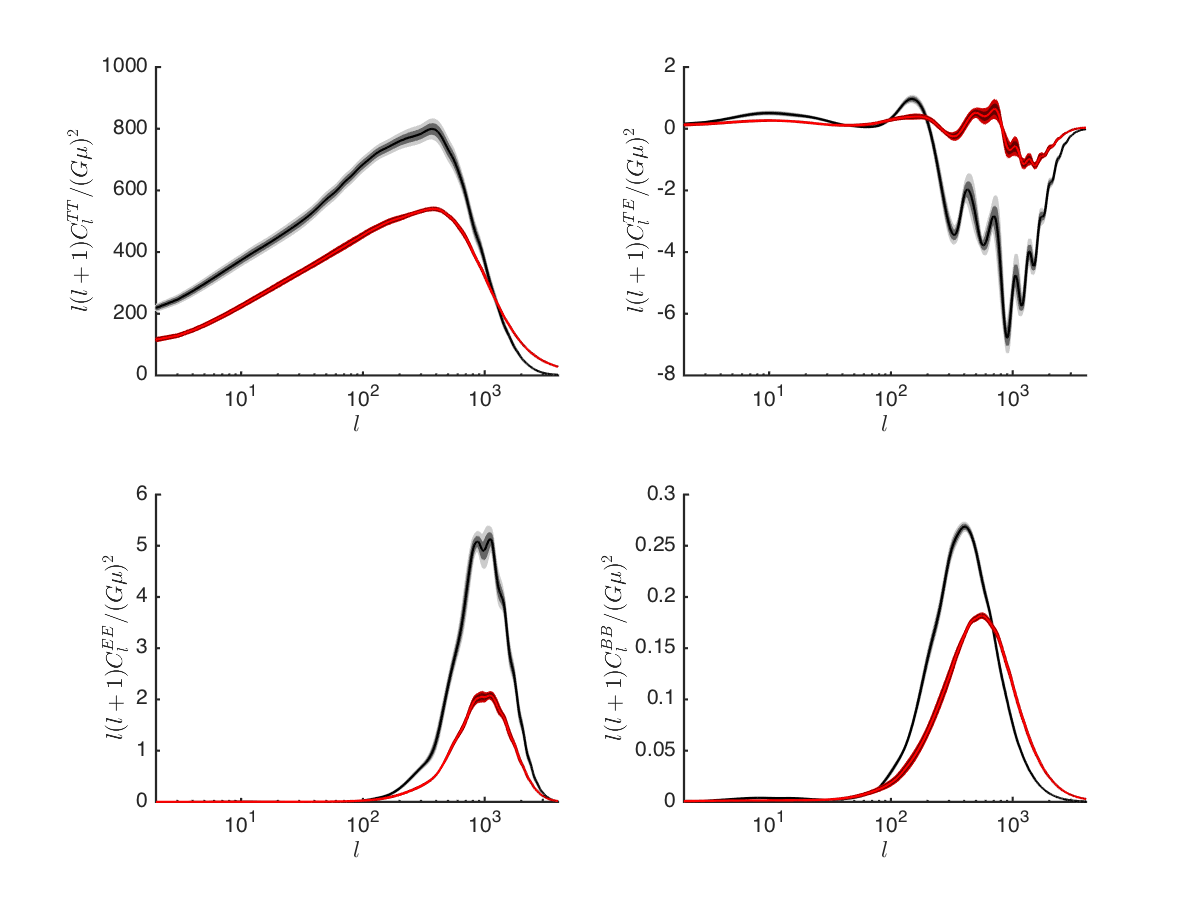} 
    \caption{Temperature and all polarization channels for the CMB of O($2$) (black line) and Abelian Higgs (red line).  Solid lines 
    correspond to the mean spectra while shaded regions represents $1\sigma$ and $2\sigma$ confidence limits obtained by bootstrapping 10 times over 5 radiation and 5 matter realizations for UETCs (over 7 radiation and 7 matter realizations in the UETC merging process for AH) \cite{Lizarraga:2016onn}. Note that $\mu = \pi \eta^2$, where $\eta$ is the vacuum expectation value of the scalar fields. 
    } 
    \label{fig:cl-com}
 \end{figure}

\begin{figure}[htbp]
    \centering
    \includegraphics[width=15cm]{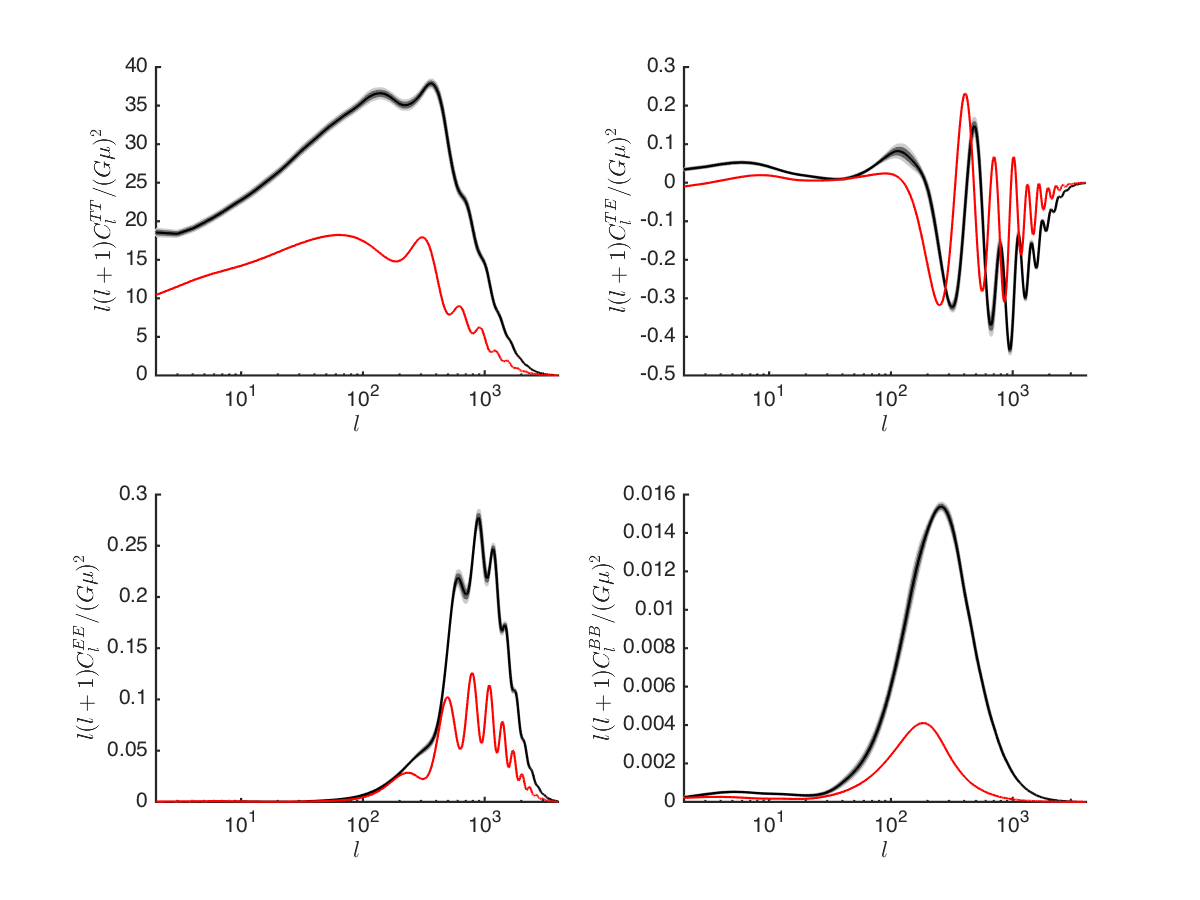} 
    \caption{Temperature and all polarization channels for the CMB of O($3$) (black line) and large-$N$ computation made in \cite{Fenu:2013tea} (red line).  Black lines correspond to the mean spectra while grey regions represents $1\sigma$ and $2\sigma$ confidence limits obtained by bootstrapping 10 times over 5 radiation and 5 matter realizations for UETCs. 
    Note that $\mu = \pi \eta^2$, where $\eta$ is the vacuum expectation value of the scalar fields.
}
    \label{fig:spc-o3}
 \end{figure}

Figs.~\ref{fig:cl-com} and \ref{fig:spc-o3}
shows the temperature and   polarization power spectra obtained for the two model as solid black lines, 
normalised using the parameter 
\be
\mu=\pi \eta^2,
\label{eq:mu}
\ee
where $\eta$ is the vacuum expectation value of the (real) scalar fields. 
In each case $\mu$ has a different meaning. For the global string case it can be seen as the tension in the core of the string,
and for global monopoles the energy in the monopole core is approximately $\mu\delta$.

In Fig.~\ref{fig:cl-com} we have also plotted the power spectra obtained for Abelian Higgs strings in \cite{Lizarraga:2016onn} as red lines for comparison, for which $\mu$ is precisely the string tension. 
The figure shows that the amplitude of the signal of global strings is almost a factor of two bigger, whereas the shape of both around the peak are very similar. The global string power spectra fall off faster at high multipole, a consequence of the faster fall-off of the ETCs at high wavenumber.

Fig.~\ref{fig:spc-o3} in turn shows the temperature and all polarization  power spectra obtained for global monopoles. 
Comparing with the O($2$) case we can see that the signal given by the O($2$) model is much bigger than the one given by O($3$) monopoles. Furthermore, although the overall shape is similar in the both cases, the O($3$) case is more oscillatory.

The power spectra for global strings and monopoles can be compared with that obtained in \cite{Fenu:2013tea} for O($N$) defects, in the large-$N$ limit (red line in Fig.~\ref{fig:spc-o3}). It can be noted that all spectra share a similar overall shape. 
The spectrum obtained from the large-$N$ limit shows clearer oscillations, 
more closely resembling the global monopole curve,   
but underestimates the amplitude. 
To quantify the underestimate, we show in table~\ref{t:largencom} the values of the power spectra for the two cases (obtained in our analysis and using large-$N$) at $l=10$ and at the peak of the power spectra. 
Note also that the ratio is not the same in both points showing that the large-$N$ limit does not capture well the detailed shape of the power spectra. In these values we can see a similar behaviour to the one shown in \cite{Figueroa:2012kw}; that is for bigger values of $N$ the ratio between the measured value and the theoretical one seems to reaching one.

\begin{figure}[htbp]
    \centering
    \includegraphics[width=7cm]{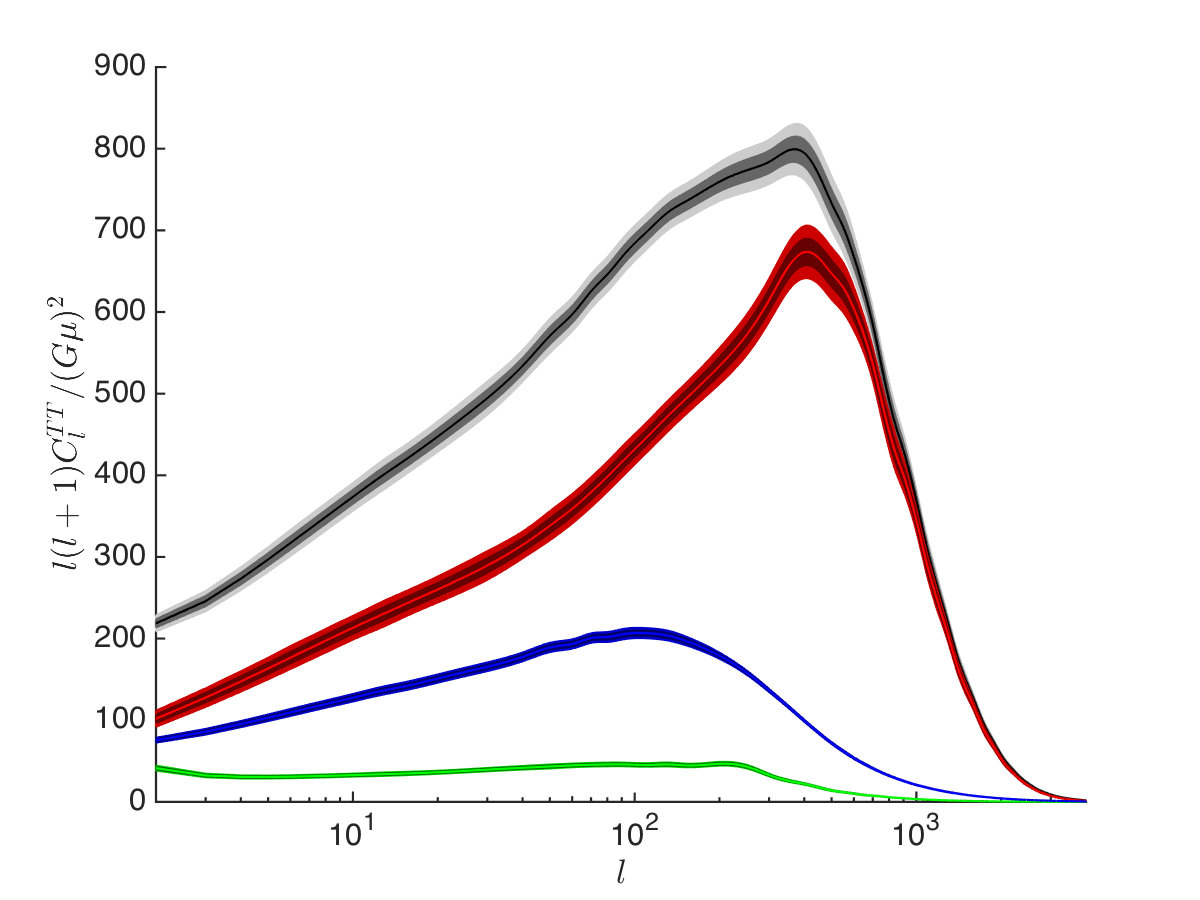} 
    \includegraphics[width=7cm]{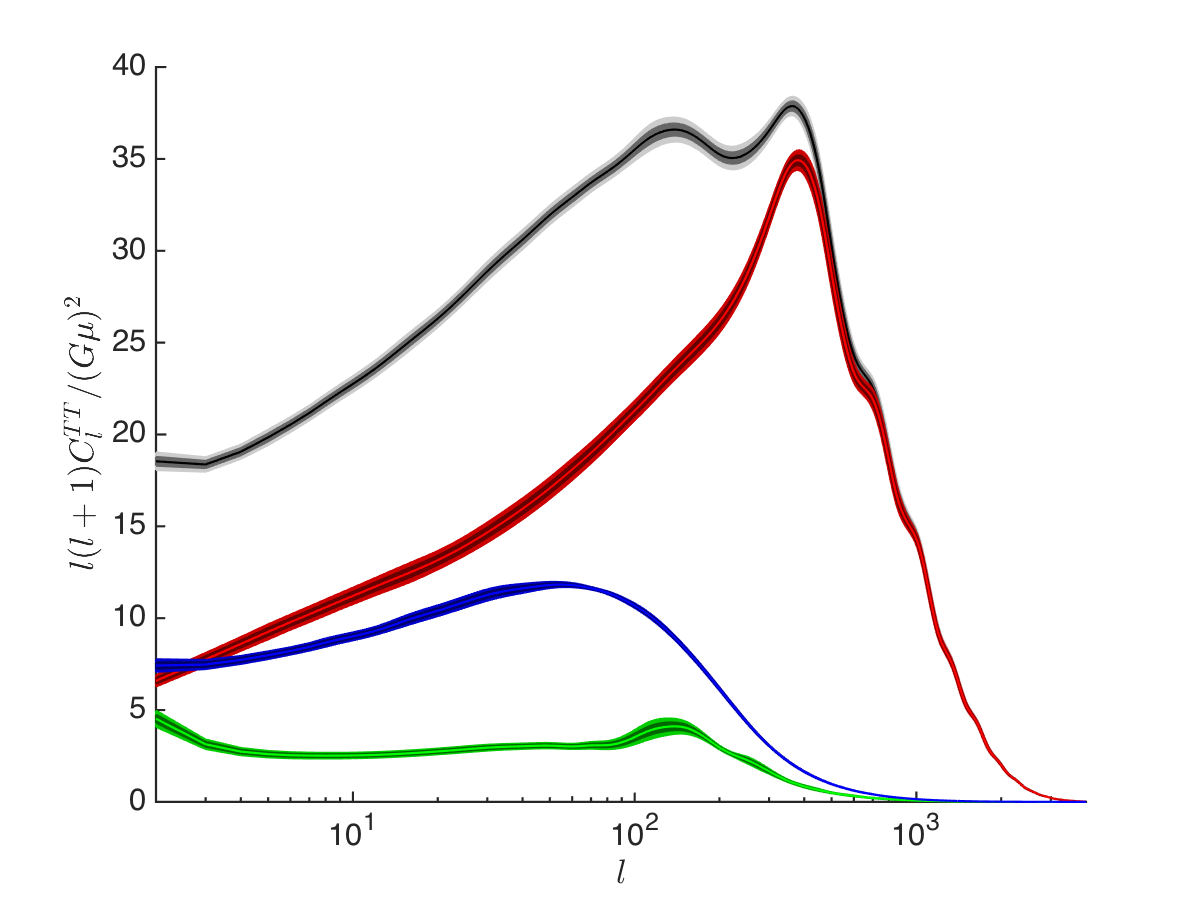} 
    \caption{The CMB temperature power spectrum for O($2$) (left) and O($3$) (right). The plot shows the total (black region) plus the decomposition into scalar (red region), vector (blue region) and tensor (green region). In those regions the bright lines correspond to mean spectra while the pale regions represents  $1\sigma$ and $2\sigma$ confidence limits obtained by bootstrapping 10 times over 5 radiation and 5 matter realizations for UETCs. }
    \label{fig:tt-all}
 \end{figure}

 Fig.~\ref{fig:tt-all} shows the contribution of scalars, vectors and tensors to the temperature channel for the both cases O($2$) and O($3$).  In this figure we can see that the contribution of scalars is the dominant one and that vectors and tensors contribute fewer to the temperature channel, being the tensor the one that contribute the least. The contribution scheme in both models, O($2$) and O($3$), is almost the same.
 
\begin{table}
\begin{center}
\renewcommand{\arraystretch}{1.2}
\begin{tabular}{|c|ccc|ccc|} 
\hline
 &  \multicolumn{3}{c|}{Peak} &  \multicolumn{3}{c|}{$l=10$}  \\
\hline
$N$       & LN & S & Ratio (S/LN) & LN & S & Ratio (S/LN) \\
\hline
2 & 41.0 & 799 & 19.5 & 32.0 & 374 & 11.7 \\
\hline
3 & 18.2 & 37.9 & 2.08 & 14.2 & 22.8 & 1.60 \\
\hline
 \end{tabular} \\ 
  \caption{\label{t:largencom} Values of $l(l+1)C_l^{TT}$ for $N=2$ and $N=3$ at the peak of the temperature power spectrum 
  and at multipole $l=10$.
  S denotes values obtained from our simulations and LN values obtained in \cite{Fenu:2013tea} using the large-$N$ limit.}
\end{center} 
\end{table}

We can also compare to the numerical simulations in the O(2) and O(3) NLSM \cite{Pen:1993nx,Pen:1997ae}. 
Although these calculations were done in an Einstein-de Sitter universe with $h=0.5$, there is little difference 
in the power spectrum with $\Lambda$CDM at $l = 10$ where the Sachs-Wolf effect is dominant.

The shapes of the power spectra 
\cite{Pen:1997ae} are broadly similar, although the relative contribution of the scalar is much lower in the calculation of Ref.~\cite{Pen:1997ae}. 
This has the effect of reducing the height of the peak relative to power at $l=10$,  and moving the peak from $l \simeq 300$ to $l \simeq 100$.
This is likely to be an effect of the higher matter density \cite{Hu:1995kot}.

The smaller simulation volume, $400^3$ is unlikely to be a source of difference, as scaling is rapidly reached in NLSM simulations. 
The matter-radiation transition was treated using a multistage eigenvector interpolation method, which gives very similar results to 
our fixed-$k$ interpolation method \cite{Lizarraga:2016onn}.

We compare the value of $G\mu$ required to normalise the temperature power spectrum to the COBE at angular scales of $10^\circ$ reported in Ref.~\cite{Pen:1997ae} with the value required to normalise our power spectra to \Planck\ at $l=10$ in Table \ref{t:GmuCom}, which are approximately equivalent.
At these large angular scales, the difference in the background cosmologies should not have much effect, and we 
conclude that there is a suggestion of a systematically higher temperature power spectrum for a given $G\mu$ 
in the NLSM.  This would be interesting to check, but would require a separate campaign of simulations in the NLSM. 

\begin{table}
\begin{center}
\renewcommand{\arraystretch}{1.2}
\begin{tabular}{|c|c|c|} 
\hline
$N$       & NLSM \cite{Pen:1993nx} & This work  \\
\hline
2 & $ (2.6\pm1.3) \times 10^{-6}$  &  $(1.34\pm0.02) \times 10^{-6}$ \\
\hline
3 & $ (2.3\pm0.4) \times 10^{-6}$  &  $(5.44\pm0.02) \times 10^{-6}$ \\
\hline
 \end{tabular} 
  \caption{\label{t:GmuCom} Comparison between $G\mu$ normalised to the COBE temperature fluctuations filtered at a $10^\circ$ scale 
   in the NLSM \cite{Pen:1993nx} and the approximately equivalent value obtained by normalising our power spectra to 
   \Planck\   at $l = 10$.}
\end{center} 
\end{table}


\section{Fits and constraints \label{sec:fits}}

The CMB anisotropy predictions obtained from field theoretical numerical simulations of global string and global monopoles are compared with the latest CMB data released by the \Planck\ collaboration \cite{Ade:2015xua}, in order to put limits on the allowed fraction of those defects. We consider the whole \Planck\ CMB dataset and analyzed them using the publicly available likelihoods (TT, TE, EE + lowTEB) provided by the collaboration \cite{Aghanim:2015xee}. The Monte Carlo analysis has been performed using {\sc cosmomc} \cite{Lewis:2002ah}, which uses {\sc camb} \cite{Lewis:1999bs} as the Einstein-Boltzman solver for the primordial perturbations.

The base model for the data consists firstly of the
6 parameters of the standard \LCDM\ model:
$\omega_b$, the physical baryon density; 
$\omega_c$, the physical cold dark matter density; 
$\Theta_{MC}$, the approximate ratio of the sound horizon at recombination to the angular diameter distance; 
$\tau$, the reionization optical depth to last scattering; 
$n_{\rm s}$, the spectral index of the primordial scalar perturbations, and 
$A_{\rm s}$, their amplitude.
There are also 27 nuisance parameters in the fit, relating to the experiments used in the analysis.
We call this the {\it Power-Law} model ($\mathcal{PL}$). 

In order to construct models with defects, we add to the power spectrum of the basic $\mathcal{PL}$ model the possible contribution of one or other of the global defects, parametrised by $10^{12}(\gmu)^2$ (\ref{eq:mu}) or equivalently $f_{10}$, the fraction the defects contribute to the temperature power spectrum at $l=10$. 
Shape variations of the defect power spectra for different background cosmologies are negligible, 
as their contribution is expected to be of order of about  $1\%$ of the temperature power spectrum. 
Note that $f_{10} \propto (\gmu)^2$, and that a flat prior with upper bound $10^{12}(\gmu)^2<100$ was imposed.

We find that the addition of the defects to the \PL\ model does not improve the fit to the data. Even though O($3$) are able to improve slightly the likelihood, the improvement is not significant. In all cases the scenario with no defects $\gmu=0$ is compatible with the measurements.

We therefore give the $95\%$ confidence level upper limits for the defect model parameters for O($2$) and O($3$) defects in Table~\ref{t:gmulimits2}. The  base-model parameters are consistent with the \LCDM\  \Planck\ values. 

The upper bound on $G\mu$ is derived without any extrapolation of the UETCs due to the expected logarithmic increase of the 
contribution from the cores of strings discussed at the end of Section \ref{s:Sca}, 
as we are unable to confirm the increase in the limited range of our simulations. If one assumes
that the UETCs scale by a factor of the ratio of the logarithms in our simulations and at decoupling, 
then the upper bound is reduced to 
\begin{equation}
10^{12} (G\mu)^2 < 0.031 \left(\frac{\log(m_{\rm s} a \xi)_{\tau_{\rm ref}}}{\log(m_{\rm s} a \xi)_{\tau_{\rm dec}}} \right)^2,
\end{equation}
where $\tau_{\rm dec}$ is the conformal time at decoupling. 
This gives
\begin{equation}
(G\mu)^2 < 7 \times 10^{-17}.
\end{equation}

\begin{table}
\begin{center}
\renewcommand{\arraystretch}{1.2}
\begin{tabular}{|c||c|c||c|} 
\hline
Dataset &  \multicolumn{3}{c|}{\Planck\ 2015 CMB} \\
\hline
Defect & O($2$) & O($3$) &  \\ \hline
Model & {\PLgmu} & {\PLgmu} &  \PL \\ \hline
$f_{10}$ & $<0.017$   & $<0.024$  & $-$   \\
$10^{12}(G\mu)^2$ & $<0.031$ & $<0.73$ & $-$\\ 
\hline 
$-\ln{\cal L}_\mathrm{max}$ & $6472$ & $6470$ & $6472$ \\\hline
 \end{tabular} \\ 
  \caption{\label{t:gmulimits2} 95\% upper limits for $(G\mu)^2$ and $f_{10}$ as well as best-fit likelihood values for different cosmological models for O($2$) global strings and O($3$) global monopoles, fitting for the \Planck\ 2015  TT, TE, EE and low TEB data. }
\end{center} 
\end{table}

Comparing the fits obtained here with the fits obtained in \cite{Lizarraga:2016onn}, where the Abelian Higgs case was analysed, we observe that global strings give a slightly bigger contribution at $l=10$, $f_{10}$, compared with the Abelian Higgs with the same symmetry-breaking scale, while the contribution that global monopoles give is the biggest. However, even though global monopoles slightly improve the global fit to the data, the fitting process in general does not show any significative preference for models with defects.

\section{Discussion and conclusions \label{sec:conclusions}}

In this paper we have computed for the first time the CMB power spectra for global strings and monopoles using field theory simulations 
of the linear sigma model. Previous numerical simulations used the non-linear sigma model, which does not capture the energy-momentum of the cores of the defects.

In order to obtain the power spectra, we computed the UETCs of the energy-momentum in our simulations, 
and extracted source functions for Einstein-Boltzmann solvers, using a recently-introduced method which better 
captures the effect of transition from radiation dominated cosmology to matter dominated one. 
We compared our predictions with the latest \Planck\ data \cite{Ade:2015xua} so as to put limits on models with O($2$) or O($3$) defects.

We investigated the scaling regime with simulations performed in pure radiation and matter eras, 
giving various scaling parameters 
in Table~\ref{tab:betas2} and Table~\ref{tab:betas3}. 
They are compatible with, and more accurately determined than 
previous numerical simulations of global strings   \cite{Moore:2001px,Yamaguchi:1999yp,Yamaguchi:1999dy} 
and for global monopoles \cite{Martins:2008zz,Lopez-Eiguren:2016jsy}.

The UETCs obtained in matter dominated era for global strings and global monopoles can be seen in Figure \ref{fig:uetc-nf2-rad2} and \ref{fig:uetc-nf2-rad3}. The shapes are similar, but the amplitudes for global strings are significantly higher than those for global monopoles. 
Global string UETCs are also somewhat higher than those for the Abelian Higgs strings. 
Although there are fewer global strings per horizon volume, their effective string tension is higher by 
the logarithmic enhancement discussed at the end of the last section, and the net result is 
more energy per horizon volume in the global case.

We computed the UETC interpolation function for the radiation-matter transition,  
using 5 different time ranges that covered most of the transition period. 
The effects of matter-$\Lambda$ transition are rather small \cite{Lizarraga:2016onn}, and we do not consider them. 

The form of the radiation-matter transition function (\ref{eq:ftau}) is consistent with other analyses and the value of $\gamma$ is compatible with the values found for AH strings \cite{Daverio:2015nva} and for global defects in the large-$N$ limit \cite{Fenu:2013tea}. 
However, the value of the exponent, $\kappa$, is differs from model to model, which confirms that the transition function is not universal between  defects. It would be interesting to determine whether this parameter $\kappa$ will reach the value proposed in \cite{Fenu:2013tea} when $N$ becomes large.

After obtaining the source functions that capture the radiation-matter transition we have computed the CMB power spectra. We have compared the power spectra with those computed in the non-linear sigma model both in numerical simulations at $N=2,3$ \cite{Pen:1993nx,Pen:1997ae} and 
in the large-$N$ limit \cite{Fenu:2013tea}. We also compare global strings with 
the gauge strings in the Abelian Higgs model \cite{Lizarraga:2016onn}. 

The overall shape of the CMB power spectra for the cases under study is fairly similar, although the power spectra for global monopoles show a more oscillatory behaviour than the strings, more like the spectra obtained in the large-$N$ limit of the NLSM. 
The NLSM in the large-$N$ limit underestimates the amplitude of the 
power spectra of global defects by a significant factor, up to 20 in the case of strings. 
The NSLM numerical simulations at $N=2,3$ are broadly compatible with our linear $\sigma$-model simulations, 
although detailed comparison is complicated by the different background cosmologies. 
There is some sign that the NLSM amplitude is larger for a given symmetry-breaking scale.

The signal coming from the global string case has a similar shape around the peak but a somewhat higher amplitude than the signal coming from the Abelian Higgs strings.  The fall-off of the temperature power spectrum at high multipole is faster in the global string case, which is a consequence of the faster fall-off of the ETCs with wave number, $k^{-2}$. This faster fall-off is indicative of some non-trivial energy-momentum correlations which disguise the expected $k^{-1}$ of randomly placed string-like objects \cite{Vincent:1996qr}. 
The monopole ETCs also fall as $k^{-2}$, contrasting with $k^0$ for randomly-placed point-like objects.

Finally, comparing the power spectra predictions with the latest CMB data released by the  \Planck\ collaboration \cite{Ade:2015xua} we put limits on the allowed fraction of those defects, given in Table \ref{t:gmulimits2}. 
We have seen that the global strings could give a slightly bigger contribution compared with the Abelian Higgs case, while global monopoles could give the biggest contribution between these three models, with a fractional contribution at $l=10$ of around 2.4\%. 
The limits correspond to constraints on the symmetry-breaking scale of $\eta 
< 2.9\times 10^{15}\;\textrm{GeV}$ for global strings 
($6.3 \times 10^{14}\;\textrm{GeV}$ with the logarithmic correction to the scaling UETCs)
and $\eta < 6.4\times 10^{15}\;\textrm{GeV}$ for global monopoles. 
The global string limit is relevant for the ultra-light axion scenario \cite{Marsh:2015xka}, 
provided that the strings are formed and not inflated away, 
and also that the axion mass is less than the inverse horizon size at decoupling so that the strings 
survive long enough to perturb the CMB.
The bound on the axion decay constant $f_a$ in this case is the same as the bound on $\eta$, 
in the axion mass range $m_a \lesssim 10^{-28}$ eV. 
The constraint applies even when ULAs do not comprise the dark matter.

Using these constraints we can estimate the maximum amplitude of the gravitational wave spectrum created by global strings and global monopoles following the calculations presented in \cite{Figueroa:2012kw}. 
Assuming that  the tensor UETC scales in the same way as the scalar and vector ones
with the logarithmic correction of the effective string tension, we can just insert 
the uncorrected upper limit on $(G\mu)^2$ (see Table~\ref{t:gmulimits2}) into the 
gravitational wave energy density \cite{Figueroa:2012kw}
\be
\Omega_{\rm GW}=\frac{650}{N}\Omega_{\rm rad}\left(\frac{G\mu}{\pi}\right)^2 \frac{\Omega_\text{GW}^\text{num}}{\Omega_\text{GW}^\text{th}},
\ee
where $\Omega_{\rm rad}=1.6\times 10^{-5}$ is the radiation-to-critical energy density ratio today, and $\Omega_\text{GW}^\text{num}/\Omega_\text{GW}^\text{th}$ is a numerically determined correction factor, equal to ($130,7.5$) for $N=(2,3)$ \cite{Figueroa:2012kw}.
We obtain that the upper limit for the amplitude of the GW spectrum is similar in both cases (global strings and global monopoles) at around $\Omega_{\rm GW} \lesssim 2\times 10^{-15}$. Comparing this value with the expected sensitivity curved of the gravitational wave observatory 
LISA \cite{Bartolo:2016ami,Caprini:2015zlo} it seems that the gravitational wave backgrounds created by global strings and global monopoles lie below the sensitivity window.

\acknowledgments

We thank Martin Kunz and Guy Moore for useful discussions. JL, AL-E and JU acknowledge support from the Basque Government (IT-979-16) and the Spanish Ministry MINECO  (FPA2015-64041-C2-1P).  AL-E is also supported by the Basque Government grant BFI-2012-228. 
MH (ORCID ID 0000-0002-9307-437X) acknowledges support from the Science and Technology Facilities Council 
(grant number ST/L000504/1). 
AL-E would like to thank Kari Rummukainen and the Helsinki Institute of Physics where part of this work was performed.
Our simulations made use of facilities at the i2Basque academic network, the Finnish Centre for Scientific Computing CSC, and the COSMOS Consortium supercomputer (within the DiRAC Facility jointly funded by STFC and the Large Facilities Capital Fund of BIS). 

\bibliography{cmbconstrainsts}

\end{document}